\documentstyle[psfig,mncite,epsf,subfigure,supertabular,longtable,amstex,amssymb]{mn}

\begin{document}
\title[Microlensing limits on numbers and orbits of extra-solar planets from the 1998-2000 OGLE events.]{Microlensing limits on numbers and orbits of extra-solar planets from the 1998-2000 OGLE events.}

\author[Tsapras, Horne, Kane]
{\parbox[t]{\textwidth}{ 
Yiannis~Tsapras$^{1,}$$^2$,
Keith~Horne$^{1,}$$^3$,
Stephen~Kane$^{1}$,
Richard~Carson$^{1}$\\
}\\
$^1$School of Physics and Astronomy, Univ. of St Andrews, Scotland KY16 9SS \\
$^2$School of Mathematical Sciences, Queen Mary, University of London, E1 4NS \\
$^3$Department of Astronomy, University of Texas, Austin TX 78712, USA\\
}
\date{resubmitted Jan 2003}
\maketitle

\begin{abstract}
We analyze three years (1998-2000) of OGLE observations of microlensing events to place limits on the abundance of planets with a planet-to-star mass ratio $q=10^{-3}$ at distances $\sim 1-4$AU from their host stars, i.e. `cool Jupiters'. We fit a total of 145 events using a maximum likelihood fit that adjusts 6 parameters. Each data point on the lightcurve allows us to exclude planets close to the two images of the source appearing on opposite sides of the Einstein ring of the lens star. We proceed to compute detection probability maps for each event, using $\Delta\chi^2$ threshold values of 25, 60, 100 and combine the results from all events to place global constraints. Our selection criteria returned 5 candidate events for a planet with mass ratio $q=10^{-3}$. Only two of these remained as plausible candidates after three were rejected due to poor data quality at the time of the anomalies. Our results suggest that less than 21 ($n$)\% of the lens stars have Jupiter-mass planets orbiting them at an orbital radius of $1 < a < 4$ AU. $n \le 2$ is the number of planet anomaly candidates that are actually due to planets. The datasets presented here were obtained from the DoPhot analysis of the events available at the OGLE website. The main conclusion of this work is that observing time is more efficiently allocated by observing many events with sampling intervals that produce non-overlapping detection zones than using intensive sampling on a small number of events.
\end{abstract}


\section{INTRODUCTION}
In 1995, very precise radial velocity (Doppler) measurements resulted in the first  detection of an extra-solar planet orbiting a main-sequence star \cite{Mayor95}. This discovery spawned a series of campaigns aimed at finding new exo-planets, placing limits on their abundance and eventually understanding their formation processes.

A variety of techniques is currently being used to search for planetary signatures. The list includes pulsar timing \cite{wols92}, transit searches \cite{Charbonneau00} and gravitational microlensing \cite{Pacz86,benrye96}. This paper considers the latter.

Photons emitted by a background stellar source are deflected as they come near the influence of the gravitational field of a massive foreground object which acts as a gravitational lens. This results in two images of the source, one on either side of the lens. If the source, lens and observer are perfectly aligned, we observe a ring image around the lens whose radius is defined as the Einstein ring radius and is characteristic of each lens. 

The Einstein ring radius is given by:
\begin{equation}
{\theta}_{E} = R_E/D_d = \sqrt{ \frac{D_{ds}}{D_{s}D_{d}} \frac{4GM({\theta}_{E})}{c^{2}}},
\end{equation}
where $D_{ds}$, $D_{d}$ and $D_{s}$ are the lens-source, observer-lens and observer-source distances respectively.
The Einstein ring provides a natural angular scale to describe the lensing geometry for several reasons. If multiple images of the source are produced, the typical angular separation of the images is of the order of 2${\theta}_{E}$. Furthermore, sources that are closer than about ${\theta}_{E}$ to the optical axis experience strong lensing and are therefore significantly amplified. On the other hand, sources which are located well outside this ring receive very little amplification.

In microlensing, the separation of the images created by the lensing effect ($ \theta \sim 10^{-3}$arcsec) is too small to be resolved by current telescopes and one can observe only the combined flux. The resulting lightcurve is symmetric in time with its maximum amplification $A_{0}$ at the time of closest approach $t_{0}$ between the projected position of the source on the lens plane and the lens itself. Its shape is described by:
\begin{equation}
A(u) = \frac{u^2 + 2}{u (u^2 + 4)^{1/2}},
\end{equation}
where $A(u)$ is the total amplification and $u$ is the unlensed angular separation of source and lens in units of the angular Einstein ring radius $\theta_{E}$ \cite{Pacz86}:
\begin{equation}
u = \frac{\theta}{\theta_{E}} = \left[ u_{0}^2 + \left( \frac{2(t - t_{0})}{t_E}\right) \right]^{1/2},
\end{equation}
where $t_E=2\theta_{E} D_d/v_{\perp}$ and $v_{\perp}$ is the transverse velocity between the source and lens.

This simple point-source point-lens (PSPL) model can describe most of the cases well enough. However, there are cases when the shape of the lightcurve is not symmetric and exhibits significant deviations. These so-called anomalies of the lightcurve can be due to several factors and have been extensively examined in recent literature \cite{Dominik99b,Wozniak97,Buchalter97,Alcock95b,Gaudi97}. The most interesting of these are the anomalies which can be attributed to the binary nature of the lens. The possibility that the binary lens system may be a star-planet system of extreme mass ratio, has spawned dedicated observing campaigns to reveal such planets \cite{benrye96,gaudi00,Gould92}.

In section 2 we discuss our sample of events and fitting method. Section 3 deals with the notion of `detection zones'. These are zones on a $\Delta\chi^2$ map as a function of planet position where the presence of a planet is excluded by our analysis. In section 4 we explain our treatment of the fit residuals and discuss what threshold value we should adopt for the analysis. The best planet candidates are presented in section 5, followed by a discussion on planet detection probabilities in section 6. In section 7 we provide the theoretical background to getting limits on the number of planets per star and we conclude with a summary of the analysis in section 8.
\begin{figure*}
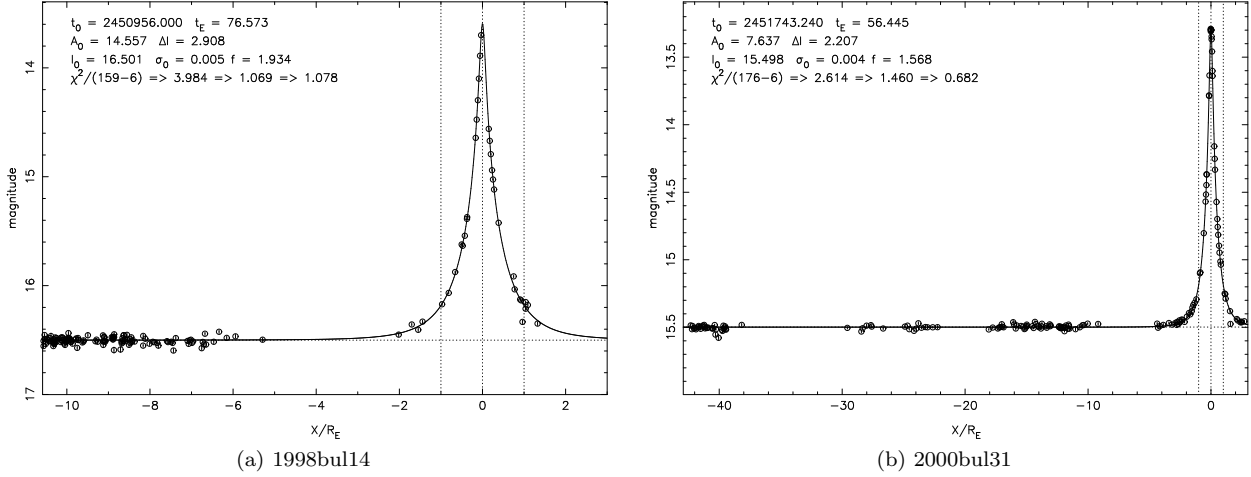

\def\subfigtopskip{4pt}
\def\subfigbottomskip{8pt}
\def\subfigcapskip{4pt}
\centering
\begin{tabular}{cc}

\subfigure[1998bul14]{\label{fig:98bul14}
\psfig{file=1998bul-14.ps,angle=270.0,width=8cm}}
&
\subfigure[2000bul31]{\label{fig:00bul31}
\psfig{file=2000bul-31.ps,angle=270.0,width=8cm}}\\
\end{tabular}
\caption{Maximum likelihood fits to the OGLE data for events 1998bul14 and 2000bul31. The fits were accomplished by adjusting 6 parameters. The best-fit parameters appear on the top left corner of the plots.}
\protect\label{fig:magplots1}
\end{figure*}
\begin{figure*}
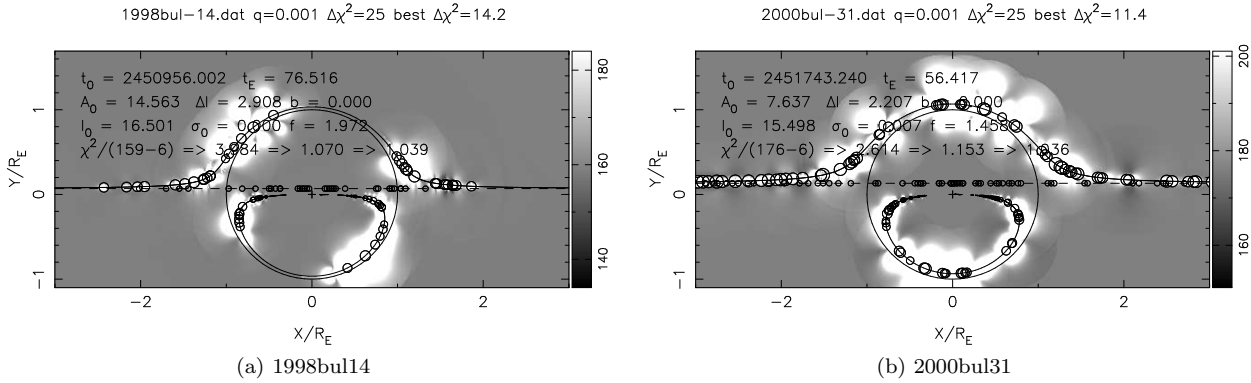

\def\subfigtopskip{4pt}
\def\subfigbottomskip{8pt}
\def\subfigcapskip{4pt}
\centering
\begin{tabular}{cc}

\subfigure[1998bul14]{\label{fig:98bul14}
\psfig{file=1998bul-14.6c.ps,angle=0.0,width=8cm}}
&
\subfigure[2000bul31]{\label{fig:00bul31}
\psfig{file=2000bul-31.6c.ps,angle=0.0,width=8cm}}\\
\end{tabular}
\caption{$\Delta\chi^{2}$ -vs- planet position for the data on events 1998bul14 and 2000bul31. The white $\Delta\chi^2 \geq 25$ detection zones show where the presence of a planet with a planet/star mass ratio $q = 10^{-3}$ is ruled out by the OGLE observations. A successful detection would have been indicated with a black zone.}
\protect\label{fig:chiplots1}
\end{figure*}

\section{ANALYSIS OF OGLE LIGHTCURVES}
\subsection{THE SAMPLE OF EVENTS}
The datasets presented here were obtained from the DoPhot analysis of the events available at the OGLE website. For the purposes of this analysis we only considered the events for which the PSPL model provided adequate fits.

Out of a total of 162 events observed by the Optical Gravitational Lensing Experiment \cite{OGLE} collaboration during a three year period (1998-2000) we found that for 145 of these the reduced $\chi^2$ of the PSPL model fit, adjusting 6 parameters, approached a value of unity indicating a good fit. We discuss our method of fitting the lightcurves in the next section.

\subsection{POINT-SOURCE POINT-LENS FITS}
In comparison with our previous work \cite{Tsapras01}, for this paper we have improved our numerical treatment of the PSPL model by replacing the approximate \cite{Gould92} formalism valid only for extreme mass ratios, $q<<1$, with a binary lens model valid for an arbitrary mass ratio.			    
Our neglect of finite-source effects \cite{benrye96} renders our PSPL lightcurves inaccurate for extreme mass ratios ($q \la u_s^2$), and high amplification ($A \ga u_s^{-1}$), where $u_s$ is the source size in units of the Einstein ring.  		    
In addition, for the discussion in Section 2.4 we allow	the lensed source to be blended with a constant source,	$A(t) \rightarrow A(t) + b$, with $b>0$, so that $f_b \equiv b/(1+b)$, with $0<f_b<1$, is the fraction of the apparent unlensed source flux contributed by nearby constant sources.
	
To illustrate our analysis we present our fits for events 1998bul14 and 2000bul31 in Figure~\ref{fig:magplots1}. These are high magnification events and the fitted event parameters are indicated on the top left corners of the plots. Our estimates of the parameters for the event 1998bul14 are in agreement with those published by the PLANET collaboration \cite{albrow00lim}.

Our initial fits were accomplished by adjusting four parameters describing the shape of the PSPL lightcurve, the time of maximum amplification $t_{0}$, event timescale $t_{E}$ (time to cross Einstein ring diameter), maximum amplification $A_{0}$ and the baseline magnitude $I_{0}$. The residuals of the 4-parameter fits were generally consistent with the estimated error bars during the bright phases, but often larger than expected near the baseline magnitude $I_{0}$. This probably reflects the difficulty in obtaining accurate photometric measurements in crowded fields (see section 2.3).

To assess evidence of lightcurve deviations at a first glance, we impose a criterion of $\chi^2/(N-4) < 3.5$, for the 4-parameter fit, which helps to identify the obvious non-PSPL lightcurves ($N$ is the number of data points fitted). This criterion was not met by 46 events. Upon further examination of these lightcurves, we concluded that  for 17 of these, the effects of stellar binary lensing, lensing of binary sources, parallax motion and stellar variability were responsible for anomalies in the lightcurve (see appendix A). We removed these 17 events from our list since our model does not account for these effects. The remaining 29 events were kept for further analysis despite the inferior quality of the lightcurves. These were events that were either sampled only during one part of the amplification phase, were imaged very sparsely for the duration of the amplification and/or showed significant scatter in the baseline.
\begin{figure*}
\def\subfigtopskip{4pt}
\def\subfigbottomskip{8pt}
\def\subfigcapskip{4pt}
\centering
\begin{tabular}{cc}

\subfigure[Blend fraction $f_b$-vs-$A_0(7)/A_0(6)$]{\label{fig:b0va0}
\psfig{file=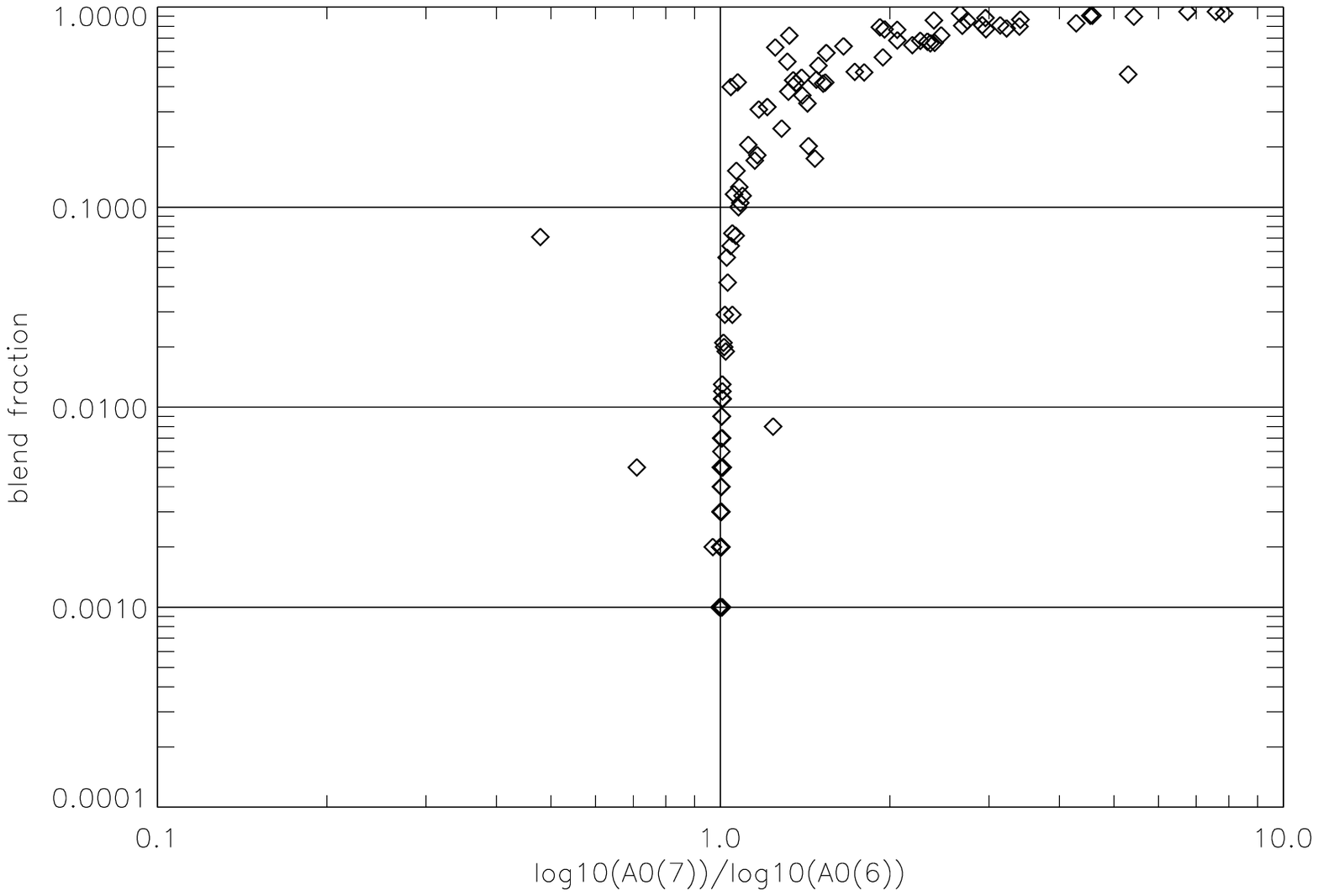,angle=0.,width=8cm}}
&
\subfigure[Blend fraction $f_b$-vs-$t_E(7)/t_E(6)$]{\label{fig:b0vte}
\psfig{file=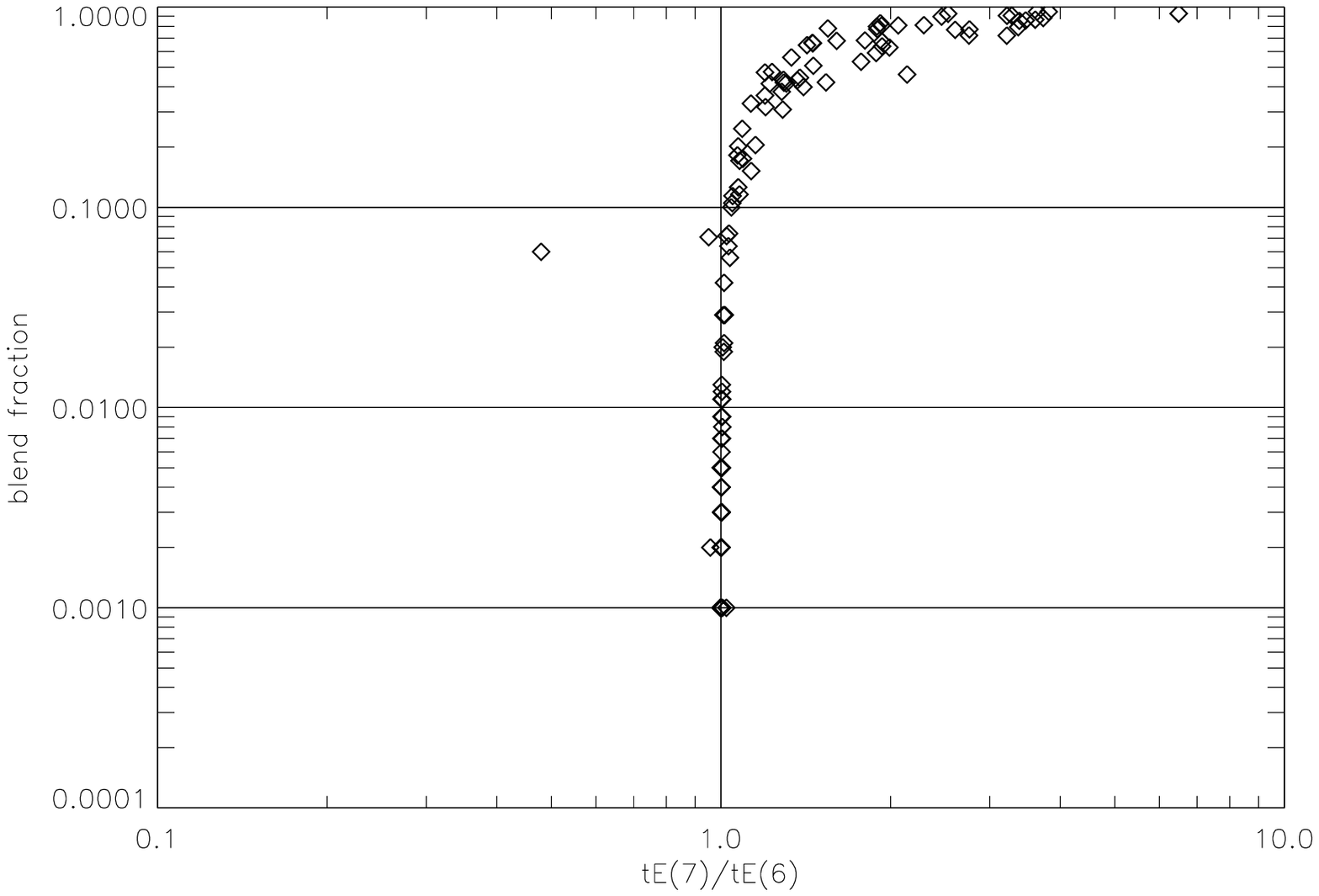,angle=0.,width=8cm}}\\
\end{tabular}
\caption[$A_{0}$ and $t_{E}$ dependence on the blend fraction for all 145 events.]{$A_{0}$ and $t_{E}$ dependence on the blend fraction for all 145 events. Allowing for a blending parameter in the fits increases both the maximum amplification and the event timescale for events with significant blending ($f_b>10$\%).}
\protect\label{fig:blenddep}
\end{figure*}

\subsection{ADJUSTMENT OF THE ERROR BARS}
Since our assessment of the evidence for lensing by planets is based on the significance of the residuals to the fitted PSPL lightcurve, we cannot ignore the systematic discrepancy between the fit residuals and the published error bars. When we simply scale the reported error-bars by a factor $f$, we find that the residuals are often larger than expected in the flat part of the lightcurve, and smaller than expected in the amplified parts of the lightcurve. This suggests that an additive flux error, in addition to an overall scaling of the error bars, may be needed to account for the rms residuals in all parts of the lightcurve. To deal with this we decided to include a crowded field error in our model, $\sigma_{0}$. This parameter accounts for increased scatter in the measurements of the unlensed flux. Thus, if blending and variable seeing effects cause the baseline flux to vary by more than the theoretical limit from photon-counting noise due to the background photons, the increase in $\sigma_{0}$ allows for that extra variance. We also scale the reported error bars by a factor $f$. Thus the error bar $s_{i}$ on the flux assumes the form:
\begin{equation}
s_{i} = \left( {\sigma_{0}}^2 + f^2 {\sigma_{i}}^2\right)^{1/2},
\end{equation}
where $\sigma_{0}$ is the additive flux error intended to account for the crowded field effects and $\sigma_{i}$ is the original error bar. The initial estimates for $\sigma_{0}$ and $f$ are 0 and 1 respectively. This 2-parameter noise model gives a more satisfactory description of the residuals for all events. 

With these two extra parameters to adjust the size of the error bars, we cannot use $\chi^2$ minimization to optimize the fit since the fit can achieve $\chi^2 = 0 $ by making the error bars infinitely large. Therefore we use a maximum likelihood criterion. Assuming Gaussian error distributions, maximizing the likelihood is equivalent to minimizing
\begin{equation}
\chi^2 + 2 \displaystyle\sum_{i=1}^{N} {\mbox{ln}}(s_{i}).
\end{equation}
This introduces an appropriate penalty for making the error bars large since that increases the value of the second term in the equation.
\begin{figure*}
\def\subfigtopskip{4pt}
\def\subfigbottomskip{8pt}
\def\subfigcapskip{4pt}
\centering
\begin{tabular}{cc}

\subfigure[$A_0$(PSF)(6) vs $A_0$(DIA)]{\label{dia1}
\psfig{file=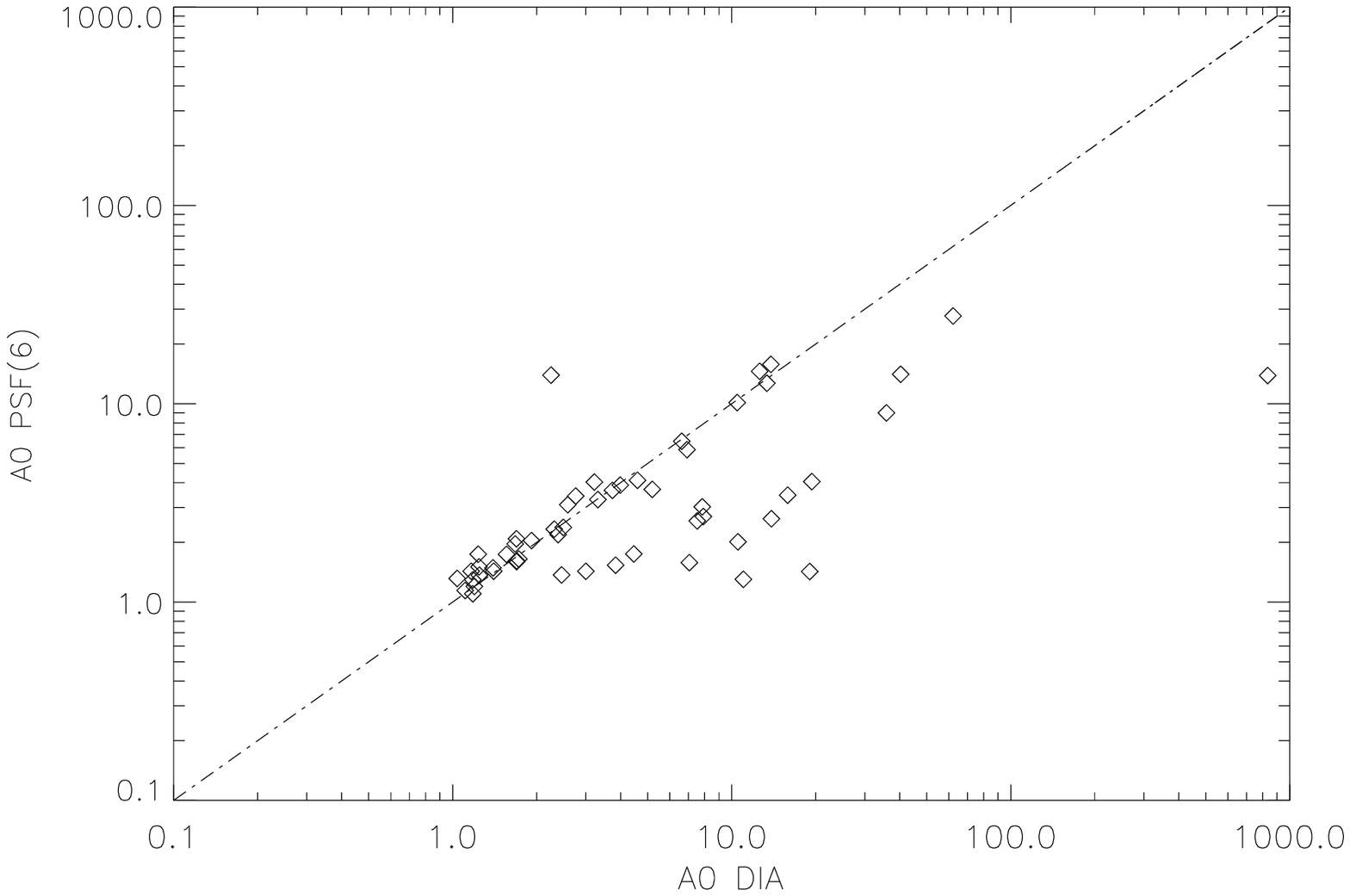,angle=0.,width=8cm}}
&
\subfigure[$A_0$(PSF)(6) vs $A_0$(PSF)(7)]{\label{dia2}
\psfig{file=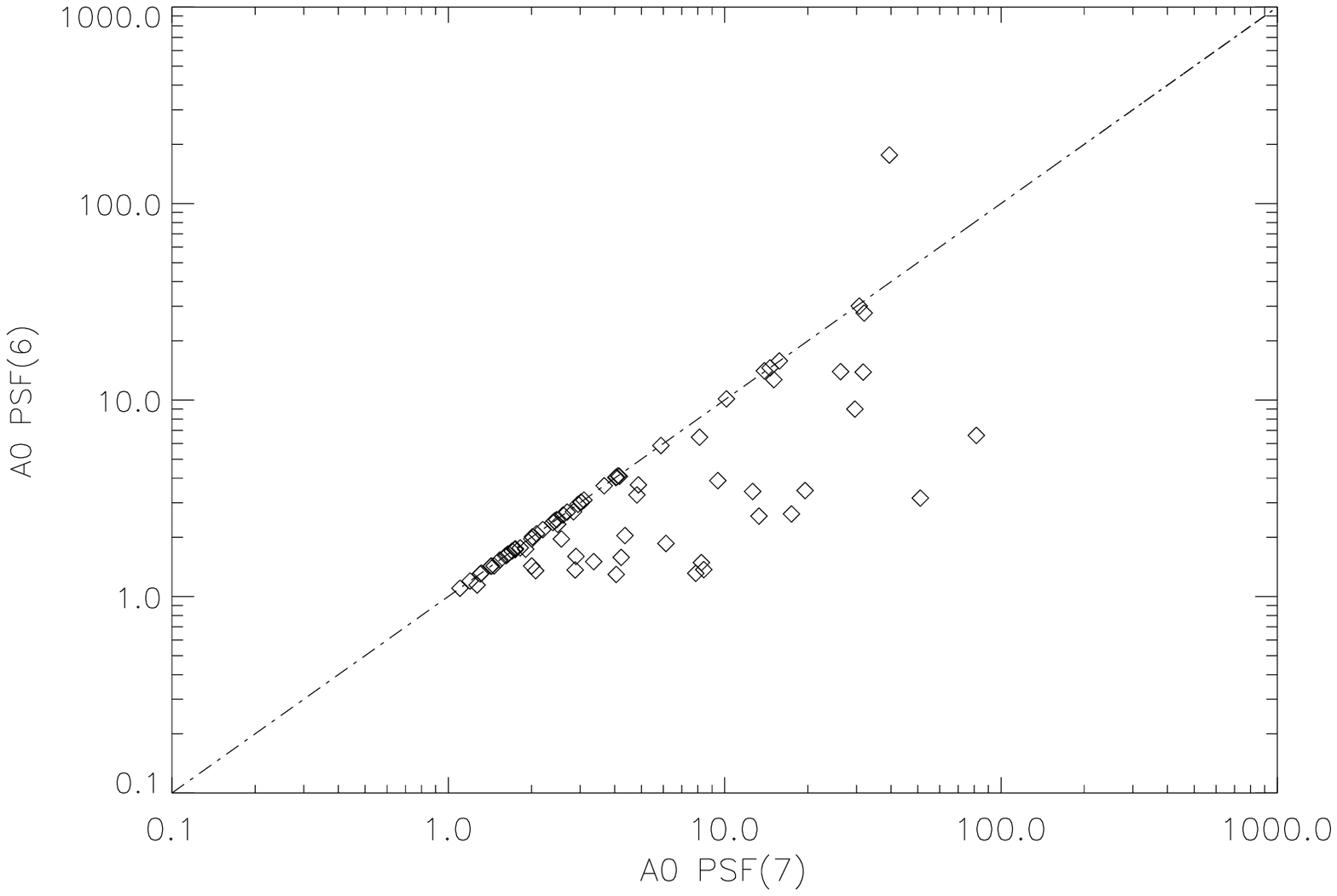,angle=0.,width=8cm}}\\
\end{tabular}
\caption[$A_0$(PSF)(6) vs $A_0$(DIA)]{\small Comparison (fig~\ref{dia1}) of the derived values for the maximum amplification from our 6 parameter fits to the OGLE data with the maximum amplification derived by difference image analysis as presented in \scite{wozniak01}. Comparison of the maximum amplifications returned by the 6 and 7 parameter fits (fig~\ref{dia2}).}
\protect\label{dia}
\end{figure*}

{\scriptsize{
\begin{table*}
\centering
\caption{Best-fit parameters for the 1998 OGLE events}
\protect\label{tab:1998}
\vspace{5mm}
\begin{tabular}{lccccccccccccl}
\hline
 Event & $t_0$ (245+)& $t_E$ & $I_0$ & $A_0$ & $N$ & $\frac{\chi^2}{N-4}$ & $\sigma_0$ & $\frac{\chi^2}{N-5}$ & $f$ & $\frac{\chi^2}{N-6}$ & $\Delta\chi^2$ \\
 & HJD & days & mag & mag & & & mag & & & & \\
\hline
1998bul01 & 0887.323 & 77.59  & 17.181 & 1.292  & 160  & 2.01 & 0.037 & 1.02 & 0.385 & 1.04 & 26.580 \\
1998bul02 & 0890.491 & 117.50 & 18.045 & 1.078  & 63   & 2.06 & 0.067 & 1.07 & 0.637 & 1.11 & 6.055 \\
1998bul03 & 0901.161 & 91.77  & 17.034 & 0.578  & 86   & 3.58 & 0.000 & 1.12 & 1.846 & 1.08 & 7.974 \\
1998bul04 & 0913.623 & 34.67  & 17.227 & 0.732  & 96   & 2.32 & 0.000 & 1.10 & 1.490 & 1.06 & 4.002 \\
1998bul05 & 0914.159 & 38.30  & 18.337 & 2.049  & 96   & 2.68 & 0.001 & 1.10 & 1.602 & 1.07 & 9.663 \\
1998bul06 & 0915.882 & 32.66  & 15.593 & 0.388  & 158  & 1.93 & 0.000 & 1.07 & 1.372 & 1.04 & 2.931 \\
1998bul07 & 0917.095 & 52.21  & 17.483 & 1.348  & 71   & 6.80 & 0.001 & 1.11 & 2.534 & 1.09 & 6.704 \\
1998bul08 & 0923.980 & 82.27  & 17.355 & 0.518  & 156  & 0.99 & 0.018 & 0.98 & 0.775 & 1.04 & 8.101 \\
1998bul09 & 0926.613 & 62.68  & 17.259 & 0.510  & 94   & 1.76 & 0.000 & 1.13 & 1.300 & 1.07 & 14.587 \\
1998bul10 & 0927.448 & 98.26  & 18.933 & 1.474  & 82   & 0.98 & 0.011 & 0.98 & 0.956 & 1.08 & 11.813 \\
1998bul11 & 0930.839 & 20.34  & 17.674 & 1.050  & 96   & 1.46 & 0.000 & 1.39 & 1.183 & 1.07 & 15.624 \\
1998bul13 & 0944.880 & 102.09 & 17.002 & 1.227  & 122  & 1.52 & 0.026 & 0.99 & 0.161 & 1.05 & 5.735 \\
1998bul14 & 0956.002 & 76.51  & 16.501 & 2.909  & 159  & 3.98 & 0.000 & 1.07 & 1.970 & 1.04 & 28.193 \\
1998bul15 & 0944.358 & 36.30  & 18.641 & 2.855  & 83   & 2.47 & 0.000 & 1.54 & 1.534 & 1.08 & 17.236 \\
1998bul16 & 0934.260 & 31.39  & 17.821 & 0.919  & 130  & 1.73 & 0.000 & 1.16 & 1.294 & 1.05 & 1.236 \\
1998bul17 & 0949.425 & 15.12  & 16.349 & 1.420  & 71   & 3.03 & 0.001 & 1.10 & 1.690 & 1.09 & 1.666 \\
1998bul18 & 0971.075 & 13.90  & 15.463 & 1.535  & 160  & 2.96 & 0.017 & 1.03 & 0.029 & 1.04 & 4.945 \\
1998bul19 & 0965.934 & 54.94  & 18.867 & 1.519  & 94   & 1.22 & 0.069 & 1.03 & 0.912 & 1.07 & 10.475 \\
1998bul20 & 0967.167 & 18.96  & 16.819 & 0.606  & 93   & 1.39 & 0.014 & 1.05 & 0.885 & 1.07 & 3.022 \\
1998bul21 & 0992.449 & 51.30  & 15.553 & 0.977  & 71   & 7.92 & 0.000 & 1.16 & 2.733 & 1.09 & 6.269 \\
1998bul22 & 0990.462 & 13.88  & 15.972 & 2.027  & 91   & 6.14 & 0.000 & 1.13 & 2.424 & 1.07 & 2.852 \\
1998bul23 & 0997.516 & 32.59  & 15.350 & 0.602  & 82   & 16.36 & 0.000 & 1.08 & 3.946 & 1.08 & 2.259 \\
1998bul24 & 0995.343 & 27.63  & 15.992 & 0.145  & 96   & 1.92 & 0.000 & 1.12 & 1.357 & 1.07 & 0.911 \\
1998bul25 & 1041.422 & 65.74  & 17.725 & 0.775  & 125  & 1.71 & 0.001 & 1.07 & 1.287 & 1.05 & 13.926 \\
1998bul26 & 1039.446 & 42.77  & 16.742 & 0.385  & 182  & 2.21 & 0.012 & 1.04 & 1.329 & 1.03 & 10.447 \\
1998bul27 & 1048.778 & 46.16  & 14.863 & 0.197  & 167  & 2.78 & 0.012 & 1.00 & 0.574 & 1.04 & 6.638 \\
1998bul30 & 1043.447 & 25.74  & 18.801 & 1.409  & 158  & 1.74 & 0.062 & 1.03 & 1.019 & 1.04 & 12.421 \\
1998bul31 & 1061.457 & 61.52  & 16.575 & 0.389  & 174  & 3.63 & 0.000 & 1.10 & 1.882 & 1.04 & 17.489 \\
1998bul33 & 1043.216 & 36.59  & 18.468 & 1.336  & 99   & 1.30 & 0.001 & 1.07 & 1.115 & 1.06 & 11.974 \\
1998bul34 & 1073.918 & 32.64  & 16.357 & 0.733  & 141  & 2.08 & 0.012 & 1.05 & 1.131 & 1.04 & 2.315 \\
1998bul35 & 1059.522 & 16.45  & 18.564 & 2.514  & 143  & 2.01 & 0.000 & 1.1 & 1.400 & 1.04 & 2.382 \\
1998bul36 & 1083.884 & 22.65  & 17.260 & 1.252  & 96   & 2.73 & 0.038 & 1.05 & 0.014 & 1.07 & 2.503 \\
1998bul37 & 1082.566 & 36.41  & 18.582 & 1.023  & 88   & 1.09 & 0.055 & 1.01 & 0.793 & 1.07 & 5.273 \\
1998bul38 & 1093.942 & 31.83  & 14.703 & 0.389  & 185  & 6.52 & 0.023 & 1.01 & 0.060 & 1.03 & 75.667 \\
1998bul39 & 1129.224 & 75.26  & 16.856 & 2.871  & 182  & 1.51 & 0.016 & 1.03 & 0.970 & 1.04 & 6.709 \\
1998bul40 & 1166.855 & 195.41 & 17.953 & 2.861  & 101  & 3.60 & 0.096 & 1.02 & 0.162 & 1.06 & 5.917 \\
1998bul41 & 1120.785 & 57.25  & 15.474 & 0.342  & 95   & 4.94 & 0.000 & 1.08 & 2.175 & 1.07 & 4.304 \\
\hline
\end{tabular}
\end{table*}

\begin{table*}
\centering
\caption{Best-fit parameters for the 1999 OGLE events}
\protect\label{tab:1999}
\vspace{5mm}
\begin{tabular}{lccccccccccccl}
\hline
 Event & $t_0$ (245+)& $t_E$ & $I_0$ & $A_0$ & $N$ & $\frac{\chi^2}{N-4}$ & $\sigma_0$ & $\frac{\chi^2}{N-5}$ & $f$ & $\frac{\chi^2}{N-6}$ & $\Delta\chi^2$ \\
 & HJD & days & mag & mag & & & mag & & & & \\
\hline
1999bul01 & 1171.868 & 175.44 & 16.969 & 0.432 & 306 & 1.96 & 0.012 & 1.05 & 1.275 & 1.02 & 16.902 \\
1999bul02 & 1257.218 & 46.15  & 15.396 & 0.942 & 263 & 3.16 & 0.002 & 1.07 & 1.752 & 1.02 & 4.647 \\
1999bul03 & 1252.012 & 170.06 & 17.864 & 2.759 & 172 & 2.69 & 0.000 & 1.16 & 1.622 & 1.04 & 14.470 \\
1999bul04 & 1260.551 & 48.79  & 17.001 & 0.295 & 223 & 2.22 & 0.001 & 1.12 & 1.474 & 1.03 & 3.503 \\
1999bul05 & 1275.200 & 96.36  & 18.829 & 3.695 & 42 & 0.62 & 0.007 & 0.63 & 0.748 & 1.10 & 6.676 \\
1999bul06 & 1274.333 & 29.14  & 15.081 & 0.543 & 201 & 8.53 & 0.025 & 1.02 & 0.459 & 1.03 & 6.261 \\
1999bul07 & 1316.248 & 68.70  & 15.929 & 0.798 & 168 & 1.84 & 0.000 & 1.11 & 1.339 & 1.04 & 6.600 \\
1999bul08 & 1287.488 & 47.09  & 19.012 & 2.385 & 200 & 1.78 & 0.001 & 1.19 & 1.321 & 1.03 & 27.262 \\
1999bul09 & 1287.673 & 29.69  & 18.893 & 5.617 & 157 & 1.13 & 0.069 & 0.99 & 0.626 & 1.04 & 1.772 \\
1999bul10 & 1294.987 & 34.06  & 17.881 & 0.851 & 220 & 1.38 & 0.037 & 1.00 & 0.731 & 1.03 & 7.294 \\
1999bul12 & 1301.822 & 43.75  & 14.594 & 0.961 & 195 & 7.90 & 0.017 & 1.03 & 1.771 & 1.03 & 8.510 \\
1999bul13 & 1318.006 & 36.58  & 15.521 & 0.595 & 155 & 1.18 & 0.008 & 0.98 & 0.637 & 1.04 & 4.385 \\
1999bul14 & 1321.115 & 40.95  & 17.341 & 0.338 & 245 & 4.40 & 0.000 & 1.10 & 2.081 & 1.03 & 3.051 \\
1999bul15 & 1309.067 & 31.1  & 19.297 & 1.921 & 229 & 1.03 & 0.091 & 0.94 & 0.737 & 1.03 & 5.477 \\
1999bul16 & 1334.422 & 50.09  & 17.155 & 0.444 & 195 & 4.23 & 0.001 & 1.06 & 2.036 & 1.03 & 2.951 \\
1999bul18 & 1319.821 & 53.32  & 18.340 & 0.603 & 155 & 1.59 & 0.068 & 1.03 & 0.815 & 1.04 & 6.599 \\
1999bul20 & 1317.034 & 4.80   & 15.246 & 0.280 & 266 & 5.90 & 0.000 & 1.13 & 2.412 & 1.02 & 1.929 \\
1999bul21 & 1318.916 & 21.58  & 18.943 & 1.199 & 256 & 1.91 & 0.145 & 1.00 & 0.798 & 1.02 & 1.675 \\
1999bul22 & 1323.570 & 14.25  & 17.677 & 1.200 & 167 & 4.00 & 0.061 & 1.03 & 1.056 & 1.04 & 4.965 \\
1999bul24 & 1335.446 & 17.90  & 18.231 & 0.619 & 156 & 3.36 & 0.001 & 1.06 & 1.808 & 1.04 & 8.50 \\
1999bul26 & 1344.629 & 12.48  & 16.613 & 0.105 & 205 & 1.70 & 0.015 & 1.02 & 0.952 & 1.03 & 3.664 \\
1999bul27 & 1366.197 & 51.55  & 17.176 & 0.498 & 254 & 1.51 & 0.000 & 1.07 & 1.220 & 1.03 & 5.827 \\
1999bul29 & 1364.779 & 55.11  & 18.852 & 1.513 & 251 & 1.84 & 0.077 & 1.08 & 1.215 & 1.03 & 5.714 \\
1999bul30 & 1358.746 & 25.82  & 18.504 & 0.675 & 154 & 2.91 & 0.000 & 1.11 & 1.685 & 1.04 & 2.457 \\
1999bul31 & 1358.456 & 11.30  & 18.143 & 9.813 & 157 & 2.45 & 0.020 & 1.07 & 1.497 & 1.04 & 5.163 \\
1999bul33 & 1434.500 & 107.32 & 16.625 & 1.173 & 223 & 1.98 & 0.014 & 1.03 & 1.058 & 1.03 & 11.805 \\
1999bul34 & 1369.641 & 12.19  & 16.398 & 1.075 & 273 & 3.17 & 0.000 & 1.10 & 1.767 & 1.02 & 3.902 \\
1999bul35 & 1391.820 & 57.91  & 18.888 & 3.606 & 209 & 1.67 & 0.006 & 1.17 & 1.280 & 1.03 & 17.714 \\
1999bul36 & 1392.576 & 56.59  & 17.643 & 2.997 & 150 & 1.24 & 0.003 & 1.13 & 1.104 & 0.95 & 12.135 \\
1999bul37 & 1398.965 & 45.08  & 16.211 & 0.297 & 166 & 4.16 & 0.012 & 1.04 & 1.788 & 1.04 & 23.457 \\
1999bul38 & 1406.990 & 81.76  & 18.033 & 0.528 & 170 & 1.50 & 0.003 & 1.05 & 1.206 & 1.04 & 8.891 \\
1999bul39 & 1437.129 & 98.09  & 17.955 & 0.759 & 116 & 2.37 & 0.036 & 1.06 & 1.183 & 1.05 & 9.317 \\
1999bul41 & 1397.784 & 11.92  & 15.627 & 1.529 & 262 & 3.12 & 0.009 & 1.04 & 1.522 & 1.02 & 1.221 \\
1999bul43 & 1405.573 & 26.96  & 18.657 & 1.037 & 167 & 2.89 & 0.075 & 1.05 & 1.425 & 1.04 & 9.219 \\
1999bul44 & 1460.023 & 64.44  & 14.561 & 0.329 & 146 & 1.28 & 0.007 & 1.01 & 0.603 & 1.04 & 6.165 \\
1999bul45 & 1420.435 & 56.24  & 17.718 & 0.465 & 277 & 2.04 & 0.040 & 1.00 & 0.876 & 1.02 & 2.502 \\
1999bul46 & 1489.964 & 66.33  & 16.602 & 0.287 & 161 & 2.79 & 0.010 & 1.07 & 1.529 & 1.04 & 3.402 \\
\hline
\end{tabular}
\end{table*}

\begin{table*}
\centering
\caption{Best-fit parameters for the 2000 OGLE events}
\protect\label{tab:2000a}
\vspace{5mm}
\begin{tabular}{lccccccccccccl}
\hline
 Event & $t_0$ (245+)& $t_E$ & $I_0$ & $A_0$ & $N$ & $\frac{\chi^2}{N-4}$ & $\sigma_0$ & $\frac{\chi^2}{N-5}$ & $f$ & $\frac{\chi^2}{N-6}$ & $\Delta\chi^2$ \\
 & HJD & days & mag & mag & & & mag & & & & \\
\hline
2000bul01 & 1585.649 & 28.58  & 18.493 & 1.780  & 274  & 2.85 & 0.104 & 1.00 & 0.658 & 1.02 & 0.733 \\
2000bul02 & 1566.209 & 166.32 & 15.075 & 0.731  & 325  & 11.24 & 0.010 & 1.02 & 3.102 & 1.02 & 10.076 \\
2000bul03 & 1581.894 & 125.90 & 17.563 & 0.669  & 307  & 1.50 & 0.001 & 1.05 & 1.218 & 1.02 & 21.671 \\
2000bul04 & 1584.128 & 52.69  & 14.990 & 0.319  & 328  & 2.42 & 0.013 & 1.00 & 0.767 & 1.02 & 4.396 \\
2000bul05 & 1552.927 & 115.79 & 15.215 & 1.106  & 241  & 2.09 & 0.008 & 1.04 & 1.174 & 1.02 & 6.635 \\
2000bul06 & 1620.938 & 45.52  & 17.196 & 3.778  & 367  & 2.22 & 0.014 & 1.1 & 1.377 & 1.02 & 10.597 \\
2000bul07 & 1615.288 & 50.73  & 16.762 & 0.685  & 452  & 1.21 & 0.003 & 1.02 & 1.084 & 1.01 & 10.795 \\
2000bul08 & 1625.078 & 99.21  & 15.708 & 0.403  & 231  & 2.96 & 0.016 & 1.02 & 1.040 & 1.03 & 8.834 \\
2000bul09 & 1614.778 & 33.15  & 17.864 & 0.799  & 267  & 2.97 & 0.000 & 1.13 & 1.711 & 1.02 & 1.634 \\
2000bul10 & 1580.839 & 109.02 & 15.918 & 0.677  & 265  & 1.17 & 0.010 & 0.98 & 0.592 & 1.02 & 5.166 \\
2000bul11 & 1616.059 & 16.88  & 17.742 & 1.443  & 304  & 1.26 & 0.018 & 1.01 & 0.903 & 1.02 & 4.568 \\
2000bul12 & 1635.963 & 51.16  & 18.875 & 4.262  & 304  & 3.36 & 0.010 & 1.64 & 1.818 & 1.02 & 43.988 \\
2000bul13 & 1654.127 & 94.47  & 15.345 & 0.197  & 355  & 1.91 & 0.012 & 0.98 & 0.632 & 1.02 & 8.264 \\
2000bul14 & 1630.096 & 55.64  & 16.972 & 0.299  & 455  & 1.83 & 0.001 & 1.06 & 1.344 & 1.01 & 6.566 \\
2000bul15 & 1630.720 & 10.94  & 15.203 & 0.950  & 271  & 10.76 & 0.017 & 1.03 & 2.626 & 1.02 & 1.455 \\
2000bul16 & 1632.040 & 46.62  & 18.437 & 2.170  & 228  & 4.67 & 0.000 & 1.22 & 2.141 & 1.03 & 31.029 \\
2000bul17 & 1648.933 & 45.81  & 16.666 & 1.822  & 334  & 11.49 & 0.000 & 1.05 & 3.370 & 1.02 & 2.758 \\
2000bul18 & 1634.764 & 7.78   & 19.004 & 0.654  & 225  & 1.41 & 0.067 & 1.01 & 0.928 & 1.03 & 3.860 \\
2000bul19 & 1652.430 & 84.27  & 18.802 & 1.110  & 242  & 1.25 & 0.021 & 1.03 & 1.072 & 1.03 & 4.523 \\
2000bul20 & 1647.332 & 51.72  & 17.968 & 1.254  & 269  & 2.93 & 0.018 & 1.07 & 1.662 & 1.02 & 7.584 \\
2000bul21 & 1655.599 & 41.59  & 19.167 & 1.592  & 219  & 1.77 & 0.073 & 1.06 & 1.168 & 1.03 & 9.235 \\
2000bul22 & 1668.179 & 217.17 & 19.932 & 1.619  & 213  & 0.95 & 0.100 & 0.94 & 0.881 & 1.03 & 8.786 \\
2000bul23 & 1665.599 & 17.26  & 18.919 & 1.236  & 293  & 3.44 & 0.001 & 1.09 & 1.843 & 1.02 & 4.693 \\
2000bul24 & 1683.499 & 45.61  & 14.203 & 0.241  & 264  & 1.12 & 0.006 & 0.89 & 0.612 & 1.03 & 10.770 \\
2000bul25 & 1685.688 & 26.78  & 16.449 & 1.307  & 362  & 1.31 & 0.007 & 1.03 & 1.019 & 1.02 & 4.581 \\
2000bul26 & 1705.615 & 67.58  & 14.045 & 0.506  & 459  & 4.69 & 0.006 & 1.07 & 2.042 & 1.01 & 11.221 \\
2000bul27 & 1681.494 & 29.39  & 18.565 & 1.464  & 426  & 2.11 & 0.072 & 1.03 & 1.153 & 1.01 & 3.251 \\
2000bul29 & 1698.301 & 39.49  & 17.067 & 1.423  & 350  & 1.68 & 0.013 & 1.02 & 1.176 & 1.01 & 8.154 \\
2000bul30 & 1780.789 & 131.43 & 15.572 & 0.536  & 273  & 2.60 & 0.001 & 1.04 & 1.600 & 1.02 & 7.867 \\
2000bul31 & 1743.240 & 56.42  & 15.498 & 2.207  & 176  & 2.61 & 0.006 & 1.15 & 1.476 & 1.04 & 11.710 \\
2000bul32 & 1725.963 & 110.04 & 16.490 & 0.130  & 231  & 2.55 & 0.000 & 1.1 & 1.583 & 1.03 & 4.375 \\
2000bul33 & 1730.173 & 73.54  & 16.991 & 2.109  & 236  & 6.56 & 0.002 & 1.15 & 2.538 & 1.02 & 21.306 \\
2000bul34 & 1710.775 & 31.00  & 17.762 & 1.367  & 297  & 3.27 & 0.000 & 1.1 & 1.796 & 1.02 & 6.018 \\
2000bul35 & 1702.909 & 17.83  & 18.469 & 4.690  & 258  & 1.27 & 0.030 & 1.02 & 1.007 & 1.02 & 11.541 \\
2000bul36 & 1712.591 & 47.39  & 16.068 & 0.137  & 285  & 2.96 & 0.009 & 1.03 & 1.507 & 1.02 & 5.656 \\
2000bul37 & 1714.100 & 26.62  & 17.633 & 0.572  & 229  & 2.21 & 0.000 & 1.1 & 1.472 & 1.03 & 4.562 \\
2000bul39 & 1719.687 & 26.66  & 17.099 & 0.294  & 360  & 2.41 & 0.015 & 1.03 & 1.375 & 1.02 & 4.027 \\
2000bul40 & 1785.392 & 115.41 & 16.542 & 0.683  & 216  & 2.13 & 0.000 & 1.08 & 1.445 & 1.03 & 5.789 \\
\hline
\end{tabular}
\end{table*}

\begin{table*}
\centering
\caption{Best-fit parameters for the 2000 OGLE events (continued)}
\protect\label{tab:2000b}
\vspace{5mm}
\begin{tabular}{lccccccccccccl}
\hline
 Event & $t_0$ (245+)& $t_E$ & $I_0$ & $A_0$ & $N$ & $\frac{\chi^2}{N-4}$ & $\sigma_0$ & $\frac{\chi^2}{N-5}$ & $f$ & $\frac{\chi^2}{N-6}$ & $\Delta\chi^2$ \\
 & HJD & days & mag & mag & & & mag & & & & \\
\hline
2000bul41 & 1767.288 & 52.95  & 13.896 & 0.703 & 409 & 4.51 & 0.009 & 1.1 & 1.739 & 1.02 & 56.209 \\
2000bul42 & 1749.297 & 82.77  & 13.540 & 0.220 & 205 & 3.14 & 0.010 & 1.04 & 1.126 & 1.03 & 9.160 \\
2000bul44 & 1745.752 & 69.86  & 17.968 & 0.427 & 233 & 1.27 & 0.019 & 1.02 & 1.007 & 1.03 & 5.690 \\
2000bul45 & 1771.437 & 133.63 & 19.208 & 1.770 & 247 & 1.41 & 0.088 & 0.99 & 0.782 & 1.02 & 19.481 \\
2000bul47 & 1751.269 & 15.78  & 16.329 & 0.206 & 448 & 1.57 & 0.000 & 1.04 & 1.247 & 1.00 & 9.242 \\
2000bul48 & 1775.745 & 37.74  & 15.764 & 1.182 & 281 & 1.71 & 0.010 & 1.01 & 0.945 & 1.02 & 4.589 \\
2000bul49 & 1751.453 & 29.60  & 17.770 & 0.464 & 291 & 2.91 & 0.001 & 1.05 & 1.694 & 1.02 & 8.319 \\
2000bul50 & 1774.360 & 41.59  & 15.827 & 0.318 & 246 & 1.93 & 0.012 & 1.00 & 0.800 & 1.02 & 5.777 \\
2000bul51 & 1762.741 & 37.61  & 17.481 & 0.449 & 454 & 1.60 & 0.036 & 0.99 & 0.792 & 1.01 & 8.305 \\
2000bul52 & 1752.887 & 5.48   & 17.690 & 1.185 & 364 & 1.84 & 0.004 & 1.04 & 1.345 & 1.01 & 3.805 \\
2000bul53 & 1759.246 & 28.37  & 19.080 & 0.668 & 278 & 1.27 & 0.001 & 1.08 & 1.118 & 1.02 & 12.044 \\
2000bul54 & 1767.331 & 51.70  & 18.451 & 1.365 & 201 & 1.90 & 0.001 & 1.16 & 1.363 & 1.03 & 11.223 \\
2000bul55 & 1783.770 & 59.98  & 18.426 & 3.632 & 284 & 3.45 & 0.108 & 1.02 & 0.987 & 1.02 & 4.705 \\
2000bul56 & 1785.608 & 32.76  & 18.136 & 1.232 & 229 & 2.04 & 0.000 & 1.09 & 1.417 & 1.03 & 7.654 \\
2000bul57 & 1775.602 & 11.34  & 18.893 & 2.057 & 271 & 1.07 & 0.073 & 0.93 & 0.697 & 1.02 & 3.862 \\
2000bul58 & 1775.378 & 5.08   & 16.683 & 0.721 & 311 & 2.19 & 0.001 & 1.04 & 1.467 & 1.02 & 3.244 \\
2000bul59 & 1779.668 & 29.00  & 19.467 & 2.767 & 247 & 2.79 & 0.210 & 1.02 & 0.969 & 1.03 & 8.730 \\
2000bul60 & 1832.841 & 78.97  & 16.396 & 0.489 & 314 & 1.29 & 0.012 & 0.99 & 0.742 & 1.02 & 5.527 \\
2000bul61 & 1858.825 & 142.31 & 15.938 & 0.278 & 238 & 2.17 & 0.000 & 1.09 & 1.460 & 1.03 & 7.539 \\
2000bul62 & 1799.773 & 42.37  & 15.554 & 0.281 & 279 & 1.59 & 0.006 & 1.03 & 1.093 & 1.02 & 6.315 \\
2000bul63 & 1808.318 & 35.44  & 17.263 & 0.478 & 258 & 3.85 & 0.006 & 1.07 & 1.934 & 1.02 & 3.017 \\
2000bul64 & 1795.913 & 29.40  & 17.502 & 2.137 & 264 & 1.08 & 0.022 & 0.99 & 0.705 & 1.02 & 8.348 \\
2000bul65 & 1838.616 & 67.55  & 15.212 & 0.436 & 356 & 2.30 & 0.006 & 1.04 & 1.328 & 1.02 & 5.079 \\
2000bul66 & 1806.554 & 56.16  & 17.392 & 0.358 & 441 & 2.02 & 0.012 & 1.04 & 1.355 & 1.01 & 11.625 \\
2000bul67 & 1808.294 & 21.12  & 16.342 & 0.361 & 361 & 4.05 & 0.000 & 1.08 & 2.002 & 1.02 & 1.246 \\
2000bul68 & 1811.16 & 7.24   & 18.196 & 1.139 & 269 & 1.39 & 0.026 & 1.03 & 1.094 & 1.02 & 4.075 \\
2000bul69 & 1821.026 & 33.42  & 16.201 & 0.187 & 237 & 1.09 & 0.007 & 0.99 & 0.884 & 1.03 & 5.463 \\
2000bul70 & 1814.593 & 33.61  & 19.110 & 0.932 & 296 & 1.07 & 0.067 & 0.99 & 0.811 & 1.02 & 2.224 \\
2000bul71 & 1973.719 & 266.81 & 16.080 & 11.008 & 234 & 16.55 & 0.000 & 1.04 & 4.036 & 1.03 & 8.861 \\
2000bul72 & 1826.773 & 31.79  & 17.524 & 0.640 & 276 & 1.92 & 0.028 & 1.02 & 0.957 & 1.02 & 3.549 \\
2000bul73 & 1830.083 & 26.05  & 18.330 & 1.534 & 397 & 4.14 & 0.007 & 1.08 & 2.022 & 1.01 & 15.936 \\
2000bul74 & 1836.064 & 95.29  & 19.027 & 4.112 & 258 & 2.19 & 0.037 & 1.21 & 1.438 & 1.03 & 13.514 \\
2000bul75 & 1868.058 & 101.78 & 18.177 & 2.941 & 314 & 1.45 & 0.003 & 1.07 & 1.194 & 1.02 & 5.004 \\
\hline
\end{tabular}
\end{table*}
}}

Tables \ref{tab:1998}, \ref{tab:1999}, \ref{tab:2000a}, \ref{tab:2000b} summarise the best-fit parameters we obtained by maximizing the likelihood of the data over 6 parameters. There are three columns in the Tables that give the reduced $\chi^2$ values for four, five and six parameters respectively, the first being evaluated with the published error bars, the latter using the method outlined above to adjust them. The last method represents a substantial improvement with $\chi^2/(N-6) \sim 1$.

\subsection{BLENDING}
An effect that we expect to be affecting all lightcurves to some extent is blending. Blending is common in the photometry of crowded fields such as the Galactic Bulge. Its influence might lead to misinterpretation of the baseline magnitude of the source and thus inaccurate estimation of the true maximum amplification and timescale of the lensing event. As blended events may be chromatic, multi-band photometry can help estimate this effect \cite{Wozniak97,Vermaak00}. 

We have performed fits to the events using 6 (no blending) and 7 (with blending) parameters. 
We chose to use the results of the 6 parameter fits in this paper. For some individual cases we also consider the 7 parameter results (also see \scite{Tsapras01}). In a poorly sampled lightcurve, a few points high or low can make the best fit indicate substantial blending even when there is no real evidence for it. Including a blending flux in the model results in larger values for the $A_0$ and $t_E$ parameters. This effect boosts the detection probabilities. Since we do not want to artificially enhance these, we choose the 6 parameter model thereby underestimating the probability of detection. 

For all 145 events, we calculate the ratio of amplifications ($A_0(7)/A_0(6)$) and event timescales ($t_E(7)/t_E(6)$) obtained by fitting the data with (7 parameter fit) and without (6 parameter fit) allowing for a blend fraction. The results are plotted in Figures \ref{fig:b0va0} and \ref{fig:b0vte}. For high blend fractions $f_b > 10$\% the effects are dramatic on the values of $A_0$ and $t_E$. We estimate that $\sim 23$\% of our events are high amplification events ($A_0 > 5$) that suffer from serious blending ($f_b > 10$\%). By not modeling this effect we are underestimating the true values of $A_0$ and $t_E$ and thus the planet detection probability for the affected events. It follows that the limits we are quoting in section 6 for the total detection probability are {\it lower} limits. The underestimate of the detection probability by not including blending in the analysis is of the order of $\sim 13$\%. We present a selection of events fitted with and without a blend fraction in Table \ref{tab:tab1}.

Some of the events from the 1998-1999 seasons have been re-identified and re-analyzed using the method of difference image analysis (DIA) by the OGLE team \cite{wozniak01}. Figure \ref{dia1} shows how our derived values for the maximum amplification (using the 6 parameter fit) compare with the difference image analysis results. Figure \ref{dia2} compares the values for $A_0$ derived using the 6 and 7 parameter fits. Both the DIA analysis and the 7 parameter fits (including blending) return a higher value for the maximum amplification.

\begin{table*}
\centering
\caption{Serious blending can have a significant effect on the fitted amplification and event timescale. This Table presents the returned best-fit parameters for selected events fitted with and without a blending parameter ($f_b$).}
\protect\label{tab:tab1}
\vspace{5mm}
\begin{tabular}{lccccccccccccccl}
\hline
 Event & $t_0$ (245+)& $t_E$ & $I_0$ & $A_0$ & $f_b=\frac{b}{1+b}$ & $N$ & $N_{par}$ & $\frac{\chi^2}{N-4}$ & $\sigma_0$ & $\frac{\chi^2}{N-5}$ & $f$ & $\frac{\chi^2}{N-6}$ & $\Delta\chi^2$ \\
& HJD & days & mag & mag & & & & & mag & & & & \\
\hline
1998bul07 & 0917.095 & 52.21  & 17.483 & 1.348 & 0.000 & 71  & 6 & 6.80 & 0.001 & 1.11 & 2.534 & 1.09 & 6.70 \\
1998bul07 & 0918.437 & 181.36 & 17.495 & 3.228 & 0.859 & 71 & 7 & 5.75 & 0.001 & 1.11 & 2.327 & 1.09 & 6.18 \\
1999bul04 & 1260.551 & 48.79  & 17.001 & 0.295 & 0.000 & 223 & 6 & 2.22 & 0.001 & 1.12 & 1.474 & 1.03 & 3.50 \\
1999bul04 & 1260.960 & 186.22 & 17.004 & 2.239 & 0.949 & 223 & 7 & 2.14 & 0.000 & 1.12 & 1.451 & 1.03 & 3.88 \\
1999bul08 & 1287.488 & 47.09  & 19.012 & 2.385 & 0.000 & 200 & 6 & 1.78 & 0.001 & 1.19 & 1.321 & 1.03 & 27.26 \\
1999bul08 & 1287.416 & 88.78 & 19.026 & 3.678 & 0.590 & 200 & 7 & 1.62 & 0.005 & 1.12 & 1.257 & 1.03 & 11.56 \\
1999bul35 & 1391.820 & 57.91  & 18.888 & 3.606 & 0.000 & 209 & 6 & 1.67 & 0.006 & 1.17 & 1.280 & 1.03 & 17.71 \\
1999bul35 & 1392.381 & 81.1 & 18.895 & 3.762 & 0.399 & 209 & 7 & 1.60 & 0.002 & 1.15 & 1.253 & 1.03 & 31.14 \\
2000bul12 & 1635.963 & 51.16  & 18.875 & 4.262 & 0.000 & 304 & 6 & 3.36 & 0.010 & 1.64 & 1.818 & 1.02 & 43.99 \\
2000bul12 & 1635.896 & 144.94 & 18.903 & 5.503 & 0.684 & 304 & 7 & 2.09 & 0.005 & 2.06 & 1.433 & 1.01 & 29.80 \\
2000bul61 & 1858.825 & 142.31 & 15.938 & 0.278 & 0.000 & 238 & 6 & 2.17 & 0.000 & 1.09 & 1.460 & 1.03 & 7.54 \\
2000bul61 & 1847.610 & 220.26 & 15.938 & 0.896 & 0.783 & 238 & 7 & 2.15 & 0.000 & 1.08 & 1.453 & 1.02 & 6.96 \\
2000bul74 & 1836.064 & 95.29  & 19.027 & 4.112 & 0.000 & 258 & 6 & 2.19 & 0.037 & 1.21 & 1.438 & 1.03 & 13.51 \\
2000bul74 & 1835.638 & 146.37 & 19.041 & 4.414 & 0.421 & 258 & 7 & 2.14 & 0.078 & 1.20 & 1.308 & 1.02 & 11.02 \\
\hline
\end{tabular}
\end{table*}

\section{PLANET DETECTION ZONES}
Microlens event monitoring in the OGLE dataset is not consistently intensive enough to detect deviations due to small planets, having sampling gaps of many hours, occasionally days. Therefore any evidence of a planetary perturbation will be confined to very few data points. In the case that a candidate deviation is noted, the corresponding frame(s) must be thoroughly examined since cosmic ray hits affecting the CCD data might be mistaken for a planetary signature. In this situation, only a very significant departure from the PSPL lightcurve could be accepted as evidence of a planet.
  
Figure \ref{fig:chiplots1} presents a $\Delta\chi^2$ map as a function of planet position for events 1998bul14 and 2000bul31, and for a planet/star mass ratio $q=10^{-3}$. These show how the $\chi^2$ of the fit changes from the PSPL fit when we add a planet at different places on the lens plane.
The gray zones on the plots indicate regions where the $\Delta\chi^2$ is zero, i.e. placing the planet here does not affect the lightcurve in any way. White zones are negative detection zones, or exclusion zones, which show regions of $\Delta\chi^2 \geq$ $+$25. These are the regions where the planet can be excluded at the time of observation since at these places the planet is close enough to one of the images to cause a large perturbation of the lightcurve near one of the data points. 
Note that the detection zones close to the Einstein ring of the lens are larger because the planet is perturbing a more highly amplified image. The detection zones also depend on the size of the error bars, with smaller error bars resulting in larger detection zones. 

The appearance of black zones on the plots would signify that a better fit has been achieved with a planet at that position interacting with one of the two images of the source star ($\Delta\chi^2 \leq$ $-$25). The black zones are positive detection zones. If a lightcurve provides evidence for a planet, the $\Delta\chi^2$ map will have a clear black spot corresponding to the planet's position. With a poorly sampled lightcurve we expect two black spots appearing at both possible image positions since we cannot tell which image the planet is interacting with. With good sampling, a black spot would show only at the correct position of the planet at the time of observation.

We have examined such $\Delta\chi^2$ maps for all 145 events and find that for the majority of the events there is no compelling evidence for a planet. However, each data point in each lightcurve rules out the presence of a planet near the major and minor image positions. Based on this method, we quantify in section 6 our non-detection results by calculating the detection probability for each event for two different mass ratios.
\begin{figure}
\centering
\begin{tabular}{c}
\psfig{file=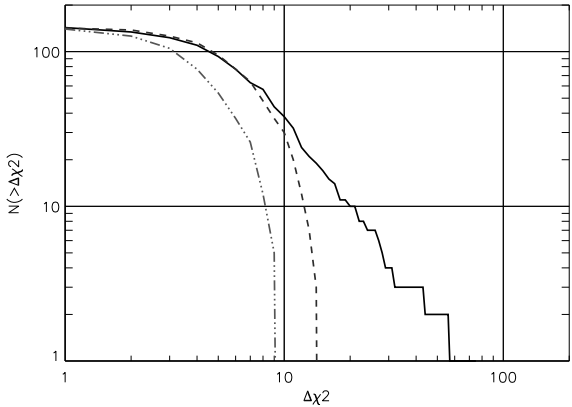,angle=0,width=8cm}
\end{tabular}
\caption[Cumulative histogram of $N(>\Delta\chi^2)$-vs-$\Delta\chi^2$]{\small Cumulative histogram of $N(>\Delta\chi^2)$-vs-$\Delta\chi^2$ that shows the total number of events that have a $\Delta\chi^2$ above any given value on the x-axis (thick line). The dashed thinner line represents the scaled expected values. The dot-dashed line represents the unscaled values.}
\protect\label{dchivn}
\end{figure}

\section{ANALYSIS OF RESIDUALS}
The 3-year OGLE dataset has 145 lightcurves with $\sim 40$ data points each (within the range $-3<a/R_E<3$) for which the PSPL model provides an adequate fit. This gives a total of $N=5840$ data points. If we consider $N$ data points with Gaussian errors, the largest residual (largest false alarm) is approximately
\begin{equation}
S_F(N) \approx 2.14 \sqrt{\log_{10}{N}},
\end{equation}
in units of $\sigma$. We therefore expect the largest single point residual in the 3-year OGLE microlensing dataset to be approximately $S_F \approx 4.15 \sigma$. If the data point with the largest residual is well separated in time from its neighbors, in comparison with the duration of the planet lens anomaly, then it should be possible for the planet lens anomaly to fit the largest residual almost exactly without being held back by other nearby data points. The largest reduction in $\chi^2$ when a planet is included in the lens model should therefore be approximately ${S_F}^2 \approx (4.15)^2 \simeq 17$.

The $\Delta\chi^2$ distribution we obtain from our fits to the 145 lightcurves with star+planet models for $q=10^{-3}$ is shown in Figure \ref{dchivn}. The median $\Delta\chi^2$ value is 6.27 and the largest is 75.67. The latter is much larger than the expected value found above. 
We make a Monte-Carlo estimate for the theoretical distribution by generating $N_{dp}$ standard Gaussian errors, where $N_{dp}$ is the number of data points in a light curve in the range $-3< x/R_E < 3$, and taking the largest outlier $S_F(N_{dp})$ for each of the lightcurves. We then plot the values of $[S_F(N_{dp})]^2$ and compare them with the observed distribution in Figure \ref{dchivn}. Although we have already corrected the error bars as described in section 2.3, the $\Delta\chi^2$ values (median 6.27) are larger than expected (median 4.18). We have stretched the theoretical distribution by a factor $f^2 = 1.5$ to match the median of the observed $\Delta\chi^2$ values. In Figure \ref{dchivn} we show the expected distribution before (dot-dashed) and after (dashed) scaling the $\Delta\chi^2$ values up. Even after this scaling, a significant tail of events with high $\Delta\chi^2$ values remains, extending up to $\Delta\chi^2$=75.

To identify events that may have anomalies due to planets, we impose a cut on events with $\Delta\chi^2 >$ 10 (last column in Tables \ref{tab:1998}, \ref{tab:1999}, \ref{tab:2000a}, \ref{tab:2000b}) in the non-Gaussian tail of Figure \ref{dchivn}. This cut leaves us with 38 events to examine. We re-fit these events including a blending parameter (7 parameter fit) and notice that for 5 of these the $\Delta\chi^2$ is now below 10 and we therefore reject them as candidates. Upon examination of this list of events we selected 5 as the most likely planetary candidates. The remaining ones were not examined any further because the deviations seemed to be due to noise or too far away from the event peak to be considered good candidates. Plots of the event lightcurves, residuals and $\Delta\chi^2$ maps for the selected events can be seen in Figures \ref{fig:devmagplots1m}, \ref{fig:devmagplots1r} and \ref{fig:00bul41m}. We briefly discuss these next.

\section{DISCUSSION OF BEST PLANET CANDIDATES}
These events were re-analyzed including a blend fraction parameter to assess the significance of the residuals.

\begin{figure*}
\def\subfigtopskip{4pt}
\def\subfigbottomskip{8pt}
\def\subfigcapskip{4pt}
\centering
\begin{tabular}{cc}

\subfigure[1998bul38]{\label{fig:98bul38m}
\psfig{file=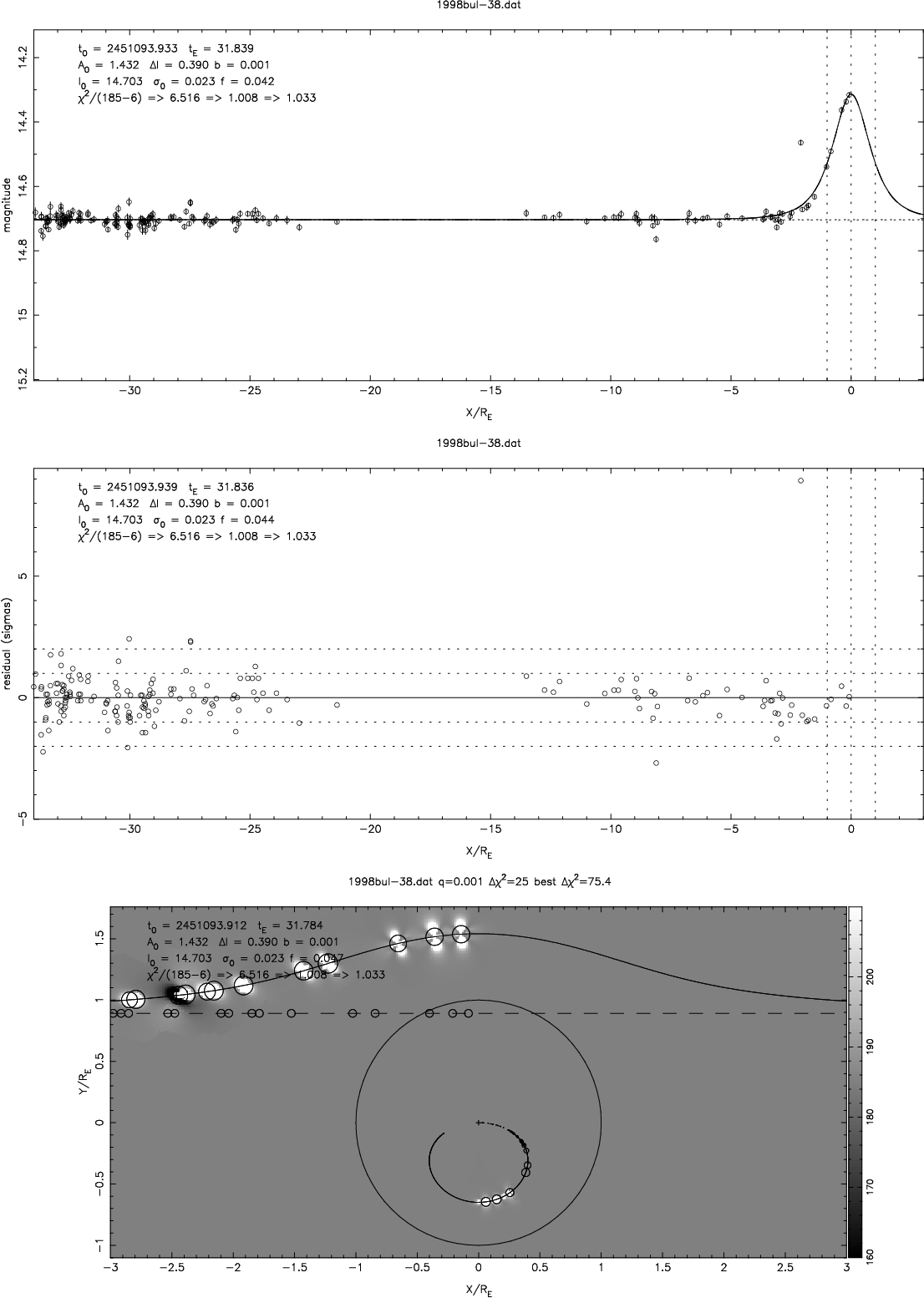,angle=0,width=8cm}}
&
\subfigure[1999bul37]{\label{fig:99bul37m}
\psfig{file=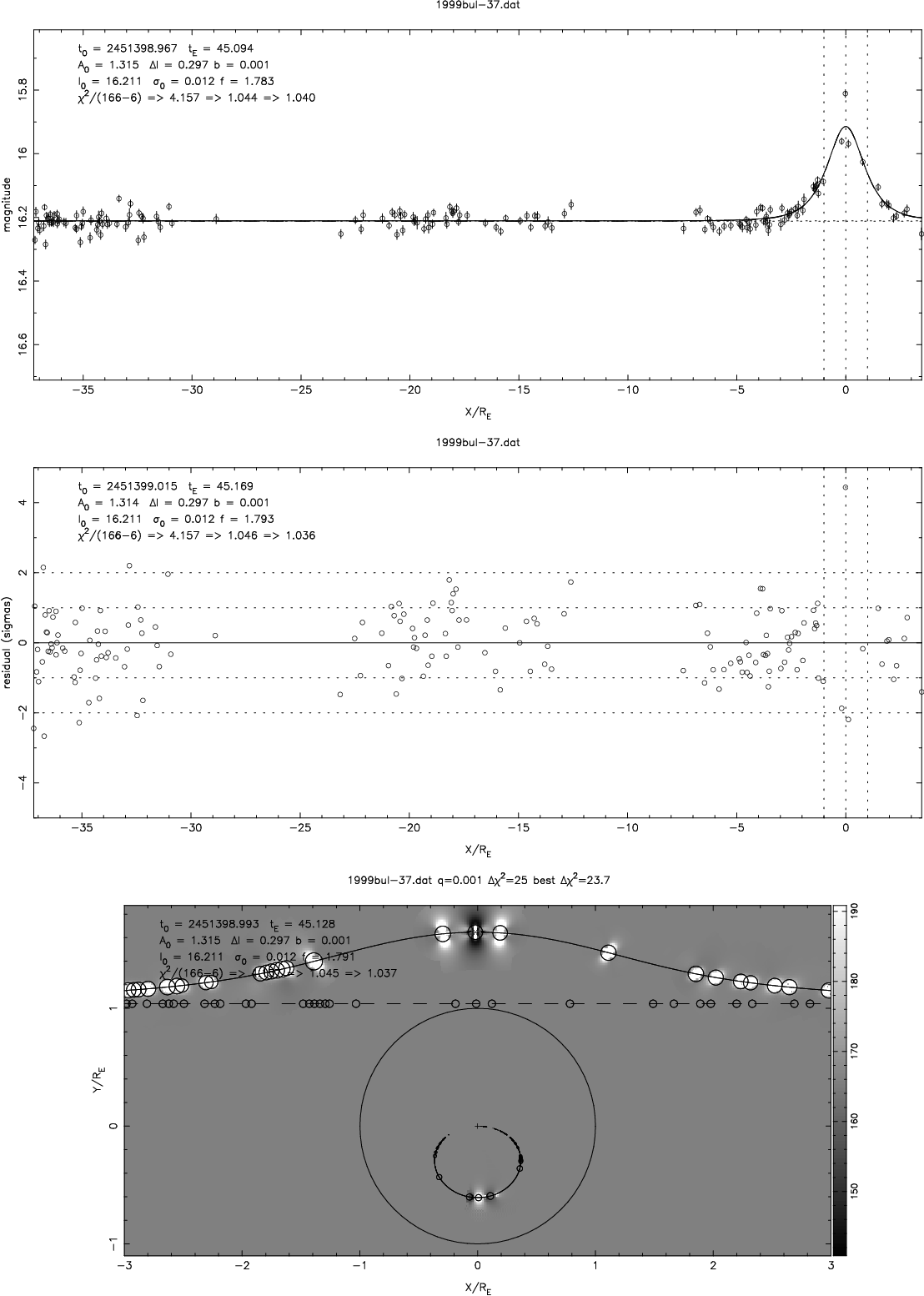,angle=0,width=8cm}}\\

\end{tabular}
\caption[Magnitude plots, normalized residuals and $\Delta\chi^2$ maps for OGLE data that deviate from the PSPL model(1).]{Magnitude plots, normalized residuals and $\Delta\chi^2$ maps for OGLE data that deviate from the PSPL model(1). The fitting for these events was done adjusting 7 parameters: ($I_0, A_0, t_E, t_0, f_b, s_0, f$). The white zones are no detection zones ($\Delta\chi^2>$25) where the planetary presence can be excluded. The gray zones are regions where there is no information about a planet and the black zones are the detection zones where there is evidence for a planet ($\Delta\chi^2<$-25).}
\protect\label{fig:devmagplots1m}
\end{figure*}

\underline{1998bul38:}
This event (Fig~\ref{fig:98bul38m}) has a baseline $I$ magnitude of 14.703 and was imaged only during the rising amplification phase. This results in a wider space for the parameters to vary, thus the constraints on such events are less stringent than on events that have been imaged on both the rise and decline of amplification. There is a single point on this lightcurve that deviates by $+9~\sigma$ at $x/R_{E} \sim -2$. OGLE has re-analyzed this event with image subtraction and the deviation is not observed anymore so we must disqualify this event as a candidate (priv. communication with A. Udalski).

\begin{figure*}
\def\subfigtopskip{4pt}
\def\subfigbottomskip{8pt}
\def\subfigcapskip{4pt}
\centering
\begin{tabular}{cc}

\subfigure[2000bul03]{\label{fig:00bul03m}
\psfig{file=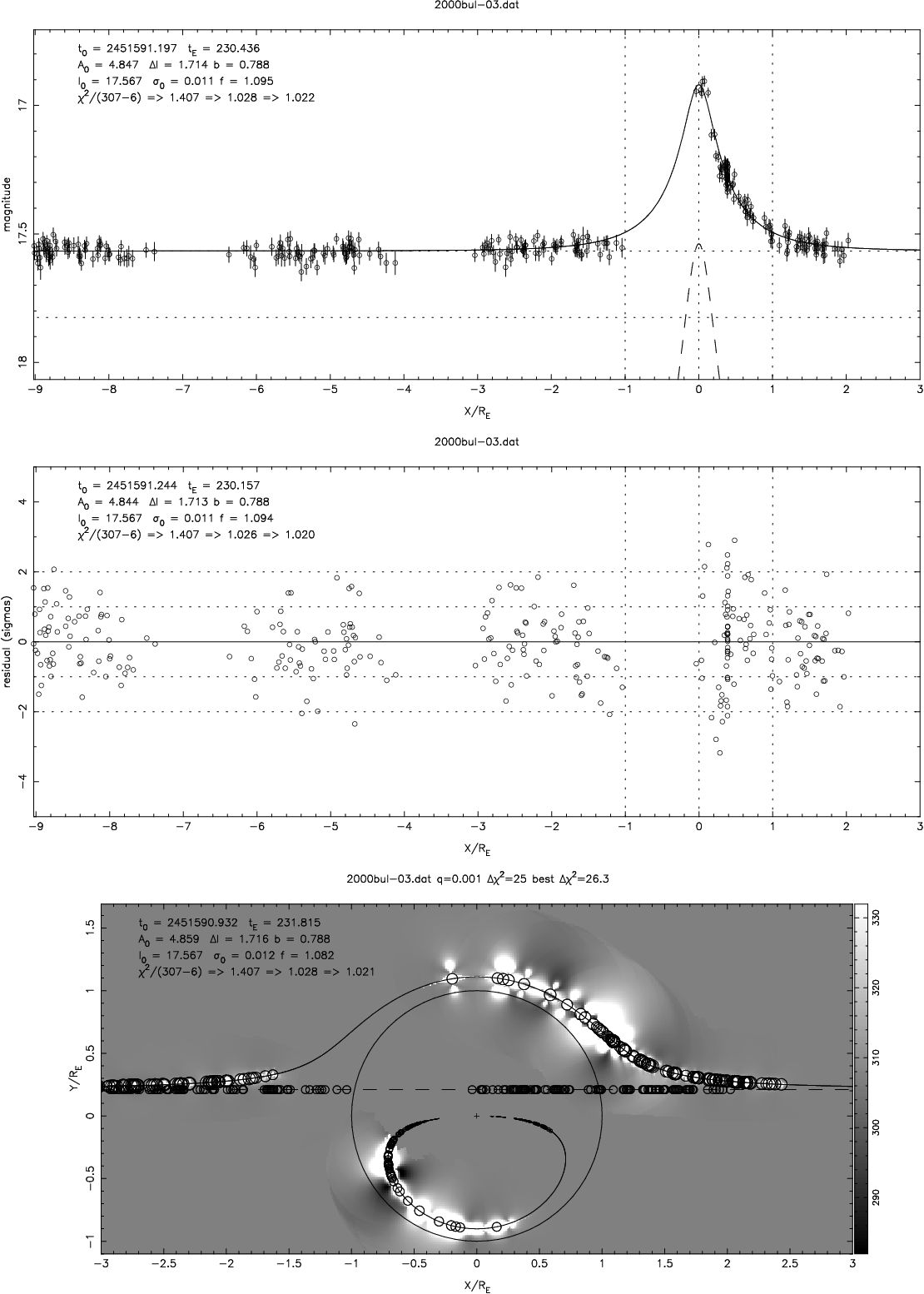,angle=0,width=8cm}}
&
\subfigure[2000bul12]{\label{fig:00bul12m}
\psfig{file=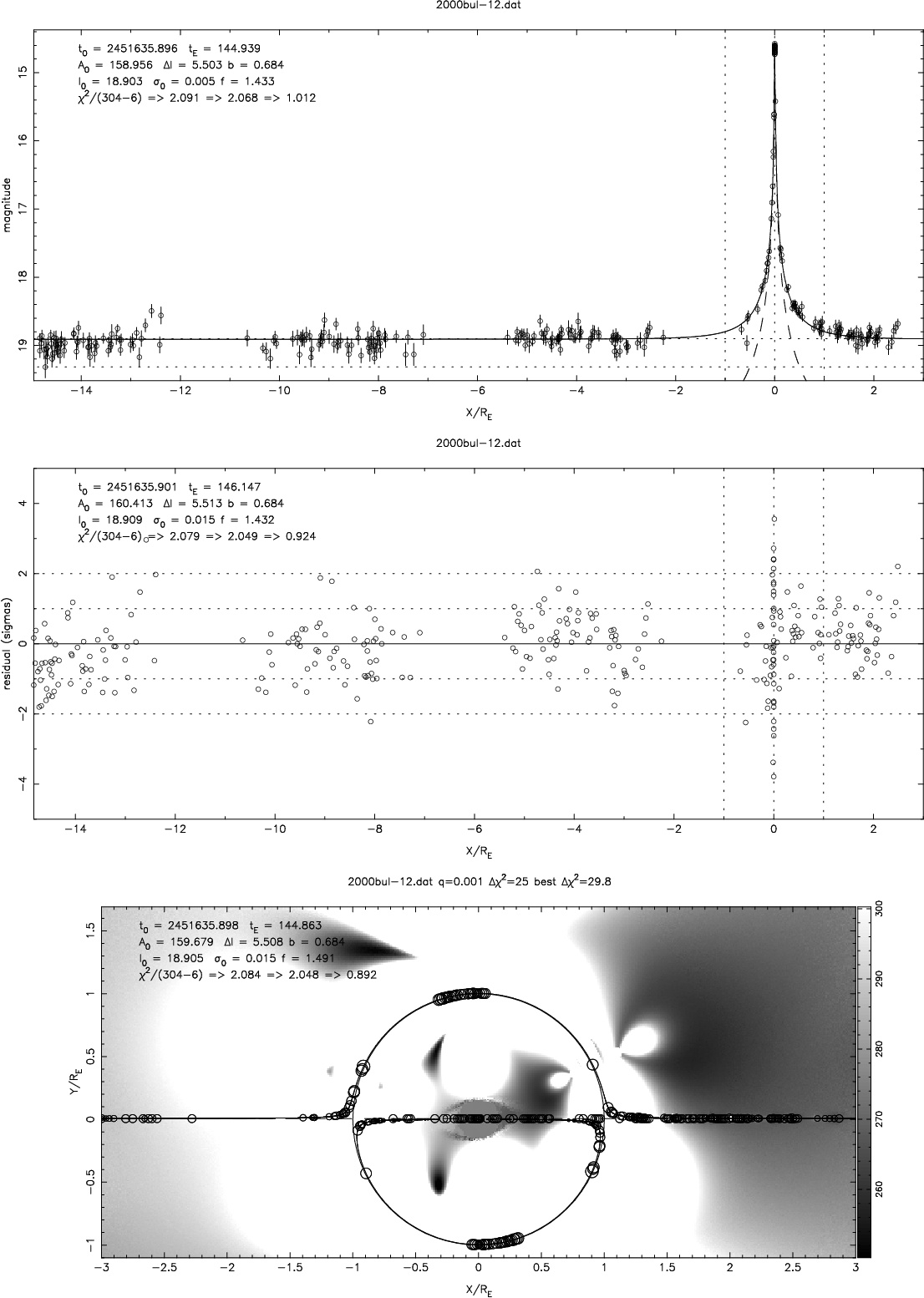,angle=0,width=8cm}}\\

\end{tabular}
\caption[Magnitude plots, normalized residuals and $\Delta\chi^2$ maps for OGLE data that deviate from the PSPL model(1).]{Magnitude plots, normalized residuals and $\Delta\chi^2$ maps for OGLE data that deviate from the PSPL model(1). The fitting for these events was done adjusting 7 parameters: ($I_0, A_0, t_E, t_0, f_b, s_0, f$).The white zones are no detection zones ($\Delta\chi^2>$25) where the planetary presence can be excluded. The gray zones are regions where there is no information about a planet and the black zones are the detection zones where there is evidence for a planet ($\Delta\chi^2<$-25).}
\protect\label{fig:devmagplots1r}
\end{figure*}

\begin{figure*}
\begin{tabular}{c}
\psfig{file=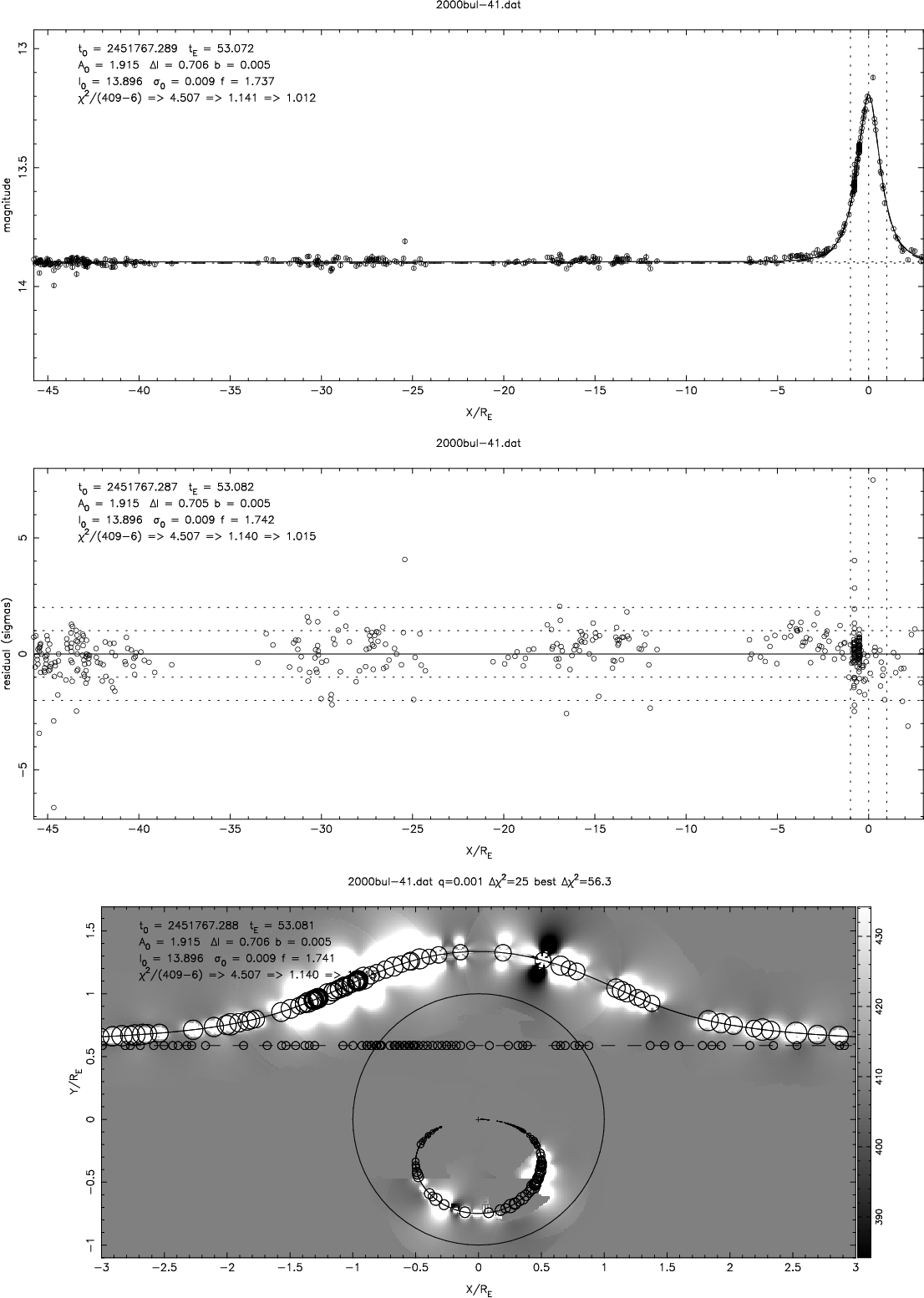,angle=0,width=8cm}
\end{tabular}
\caption[Magnitude plots, normalized residuals and $\Delta\chi^2$ maps for OGLE data that deviate from the PSPL model(1).]{Magnitude plots, normalized residuals and $\Delta\chi^2$ maps for OGLE data that deviate from the PSPL model(1). The fitting for these events was done adjusting 7 parameters: ($I_0, A_0, t_E, t_0, f_b, s_0, f$).The white zones are no detection zones ($\Delta\chi^2>$25) where the planetary presence can be excluded. The gray zones are regions where there is no information about a planet and the black zones are the detection zones where there is evidence for a planet ($\Delta\chi^2<$-25).}
\protect\label{fig:00bul41m}
\end{figure*}

\underline{1999bul37:}
Sparsely monitored about the peak, this event (Fig~\ref{fig:99bul37m}) shows a $>+4~\sigma$ residual near the peak with $-2~\sigma$ residuals on either side. The $\Delta\chi^2$ plot shows a clear dark spot. This event is disqualified as a candidate after image subtraction analysis by OGLE.

\underline{2000bul03:}
There are several data points deviating by $\sim -3~\sigma$ about the peak and the $\Delta\chi^2$ plot shows a dark spot corresponding to a deviation caused by a perturbation of the minor image. This event (Fig~\ref{fig:00bul03m}) is highly blended. It has not been imaged during the rise and there may be doubts whether this is due to microlensing.

\underline{2000bul12:}
This is a high amplification event (Fig~\ref{fig:00bul12m}) where the best PSPL model without blending fails to fit the wings and the peak data points of the lightcurve. We refit the data using our seven parameter fit which allows for a blend fraction. This fit suggests that 2000bul12 is highly blended having a blend fraction of $f_b=0.68$. Figure \ref{fig:00bul12m} shows several black zones, where the 7-parameter blend model fit is significantly improved by including a planet. We must therefore retain this event as a possible planet anomaly.

Having reanalyzed OGLE data for this event by image subtraction, the MOA team recently claimed \cite{bond01aph} that they derive a particularly high peak magnification of $\sim 160$  while their analysis raises interesting possibilities for a planetary presence. Our best-fit values for the maximum amplification are 50.7 and 159.0 for 6 and 7 parameters respectively. Note that the reported OGLE value for the amplification of this event using the standard Paczynski fit is 50.678. There is a strong correlation between the blend fraction and the derived maximum amplification which leads to uncertainty in $A_0$ for events that are significantly blended and where the blend fraction is not very well determined. HST images may be able to resolve individual sources and reduce the uncertainty.

\underline{2000bul41:}
The most interesting deviations from the PSPL model are seen in Fig~\ref{fig:00bul41m}. Several data points obtained by OGLE in the period HJD 2451745.683 to HJD 2451746.738 deviate from the fitted curve by several sigma and seem to suggest the presence of an anomaly in the region. There is also a single point close to the peak that deviates by $+7.7~\sigma$. Figure \ref{fig:00bul41m} shows a clear black zone so we retain this event as a potential planet anomaly. However, the PLANET collaboration observed this event and disregard the point close to the peak as spurious but they have obtained no data over the 2 day period before the peak. OGLE confirms that the deviation at 2451773.59746 is fully artificial since
 the image is bad due to a tracking problem. Larger than usual deviations at HJD 245174[56] are also rejected since they seem to have been caused by bad weather. This event is therefore not retained as a candidate.

To summarise the results of this section, of the 145 events                                   
which passed our original rejection criteria, 
we identify 38 events with $\Delta\chi^2 >$ 10                                 
in the non-Gaussian tail of the residuals. Of these, only 5 events show anomalous structure in the lightcurve that could plausibly be attributed to planets.

For the 5 candidates presented, none of the anomalies is well enough sampled to be securely identified as a planet. For three of these events, the causes of the deviations have been confirmed to be not of planetary nature. This leaves us with only 2 possible candidates, 2000bul03 and 2000bul12.
               
         
\section{PLANET DETECTION CAPABILITIES}
The analysis above has identified a few events with anomalies that could be due to planets. In this section  we focus our attention on the vast majority of events that are consistent with the PSPL model. The data points obtained on these events, exclude planets in certain regions near each lens. We combine the results from all the observed events to place upper limits on the number of planets per star as a function of orbit size and planet mass.
 
\subsection{PLANET EXCLUSION ZONES VS $a/R_E$ AND $q=m_p/m_\star$}
Following the lightcurve reduction and fitting, we calculate the net detection probability for each of the sampled events using the method outlined in \scite{Tsapras01}.

We calculated the probability for two different planet/star mass ratios $q=10^{-3}$ and $q=10^{-4}$ for all 145 OGLE events. Briefly, the detection probability of finding a planet at any position on the lens plane for an orbital radius $a$ is calculated as:
\begin{equation}
P(\mbox{det}|a,q) = \int P(\mbox{det}|x,y,q) P(x,y|a) dx dy.
\end{equation}
We take the first term on the right-hand side to be
\begin{equation}
P(\mbox{det}|x,y,q) = 
\begin{cases}
  1& \text{if  $\Delta\chi^{2} >$ 25} \\
  0& \text{otherwise}.  \\
\end{cases}
\end{equation}
Here, $\Delta\chi^{2}$ is the change in $\chi^{2}$ for a planet at $x,y$ relative to the no-planet model. This term becomes significant when the planet at $x,y$ lies close to one of the images of the source at the time of one of the data points in the lightcurve. The second term  $P(x,y|a)$ is obtained by randomly orienting the planet's assumed circular orbit of radius $a$ and then projecting it onto the $x,y$ plane of the sky. This term may be written as:
\begin{equation}
P(x,y|a)=\begin{cases}
\frac{\displaystyle 1}{\displaystyle 2\pi a \sqrt{a^2-r^2}} & \text{for $r=\sqrt{x^2+y^2}<a$} \\ 
0 & \text{otherwise}. \\
\end{cases}
\end{equation}
For the majority of the events the detection limits are below 5\%, while well sampled, high amplification events are dominant in imposing the most useful constraints. The total probability is given at the end of this section for detection of planets with mass ratios $q=10^{-3}$ and $q=10^{-4}$. 
Planets can be most easily detected when their orbit size                                  
matches the Einstein ring radius.                                                          
For $a<<R_E$ the planet is too close to the lens star,                                     
well inside the main detection zones that straddle the Einstein ring.                      
For large orbits,  $a>>R_E$,  the detection probability                                     
drops off as $P \propto (a/R_E)^{-2}$, equation (9),                                       
because such planets spend only this small fraction                                        
of their time projected on the sky                                                         
in the region of the star's Einstein ring.                                                     
                                                                                           
\begin{figure*}
\def\subfigtopskip{4pt}
\def\subfigbottomskip{8pt}
\def\subfigcapskip{4pt}
\centering
\begin{tabular}{cc}

\subfigure[max(P(\mbox{det} $\mid$ $a/R_{E}$))-vs-$A_{0}$]{\label{fig:pdvamax}
\psfig{file=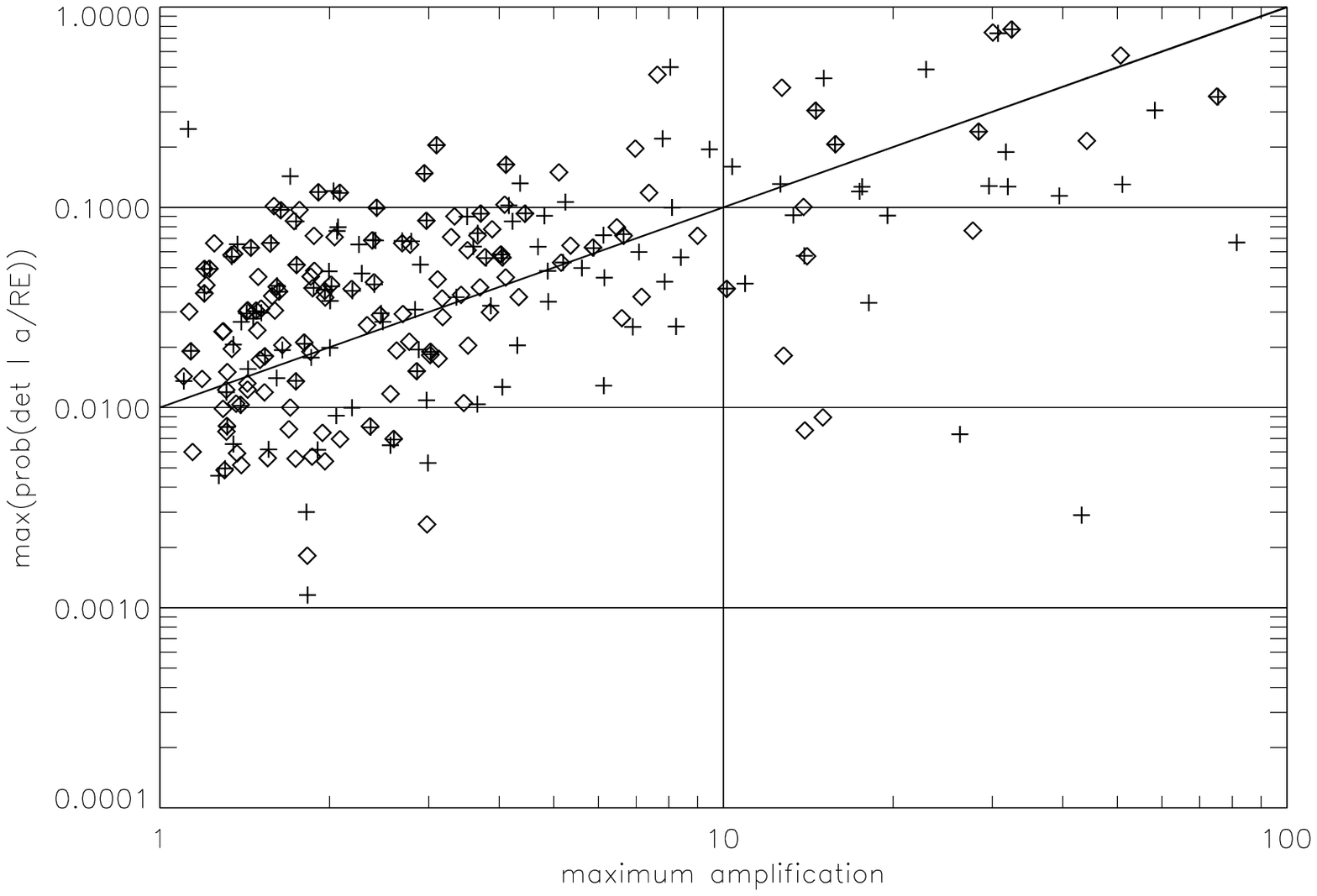,angle=0.,width=8cm}}
&
\subfigure[max(P(\mbox{det} $\mid$ $a/R_{E}$))-vs-$t_{E}$]{\label{fig:pdetvtein}
\psfig{file=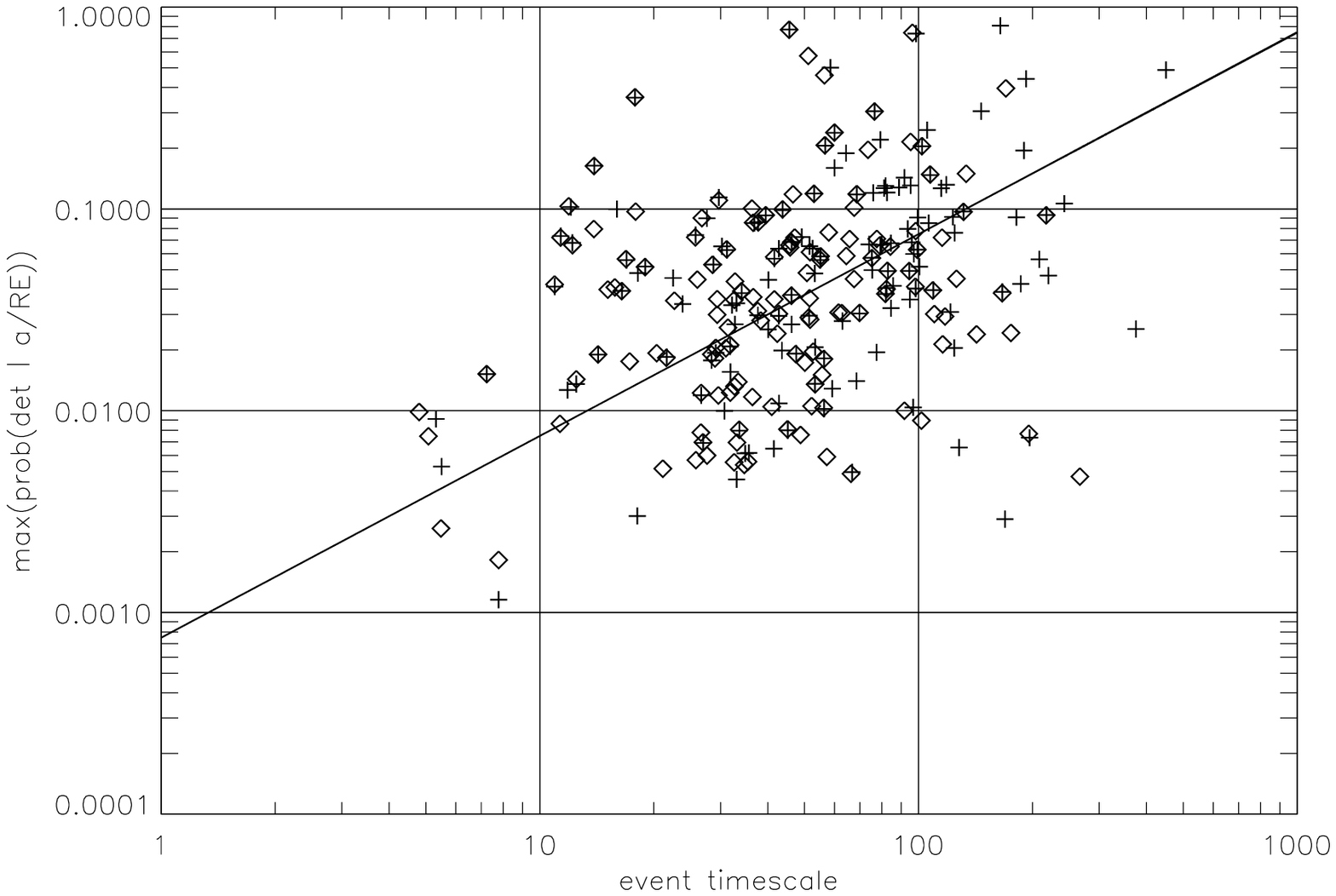,angle=0.,width=8cm}}\\
\end{tabular}
\caption{Probability of detection (for $q=10^{-3}$) dependence on $A_{0}$ and $t_{E}$ for 6 (rhombus) and 7 (cross) parameter fits.}
\protect\label{fig:pdep}
\end{figure*}

Figure \ref{fig:pdep} examines the detection probability for $a=R_E$ as a function of peak amplification $A_0$ and event timescale $t_E$. We expect the detection probability to scale with the size and number of the detection zones arising from individual measurements.                                                                              
The detection zone sizes scale roughly as the area of the planet's Einstein ring times the source amplification, while the number of detection zones scales as the event duration divided by the typical sampling time.                                                      
Thus we expect the planet detection probability from the OGLE lightcurves to scale roughly as                \begin{equation}                                                                           
P \approx 0.03                                                                     
        \left( \frac{ A_0 }{ 3 } \right)                                           
        \left( \frac{t_E}{40{\rm d}} \right)                                       
        \left( \frac{q}{10^{-3}} \right).                                          
\end{equation}                                                                             
The results in Fig.~\ref{fig:pdep} are consistent with this expectation. The strong increase in detection probability with $A_0$ is clear, while the dependence on $t_E$ is less clear. If we plot residuals relative to the above model, the remaining scatter arises from details of the actual timing of observations with respect to the lightcurve peak.

Fig \ref{fig:combpp1} shows the result of summing the detection probability $P_s(\mbox{det}|a,q)$ for star $s$ over all stars:
\begin{equation}                                                    
        P(a,q) = \sum_s P_s(\mbox{det}|a,q).
\end{equation}                                                       
This quantifies the planet detection capability of our analysis of the OGLE lightcurves. If all stars have $\eta$ planets with orbit radius $a$ and mass ratio $q$, then the number of planets we expect to detect is
\begin{equation}                                          
        <n> = \eta P(a,q).
\end{equation}
The highest detection probability occurs as expected for $a \approx R_E$. The expected $(a/R_E)^{-2}$ scaling is evident for large orbits and the probability is 10 times lower for $q=10^{-4}$ than with  $q=10^{-3}$, confirming the expected linear scaling with $q$. For comparison we also plot the curves for $q=10^{-3}$ and $q=10^{-4}$ but using a different detection threshold value of $\Delta\chi^2 \geq 60$ and 100. We see that $<n>$ scales roughly as $(\Delta\chi^2)^{-1/2}$, so that raising this $\Delta\chi^2$ threshold from 25 to 100 lowers the expected number of detections from $7\eta$ to $4\eta$.           

The detection probability curve for $q=10^{-3}$ peaks at $a/R_{E} = 1.1$ where $<n>\approx 7.4\eta$. Similarly, the curve for $q=10^{-4}$ also peaks at $a \approx R_E$ with $<n>\approx 1.3\eta$. For $\eta=1$ and $q=10^{-3}$ we expect a maximum of 7.42 detections. If the planet anomaly candidates discussed in section 5 are in fact not due to planets, $n<1$, then the number of planets per lens star is:
\begin{equation}
\eta < \frac{1}{P(a,q)} \approx 0.14 \left(\frac{q}{10^{-3}}\right)^{-1} \left(\frac{\Delta\chi^2}{25}\right)^{1/2}
\end{equation}
for planets with $a \approx R_E$.

Following \scite{gaudinab98} we can define the `lensing zone' as $a_1<a<a_2$ where $a_1=0.6R_E$ and $a_2=1.6R_E$. This is the region where detection of planets by microlensing effects is reasonably efficient. If we let each lens star have $\eta_{\rm LZ}$ planets in the lensing zone (LZ), and distribute the orbits uniformly in $\log{a}$, then the expected number of planet detections is:
\begin{equation}
\left<n\right> = \eta_{\rm LZ}
	\frac{\int_{a_1}^{a_2} P(a,q) \ d\log{a}}{\log{(a_2/a_1)}}.
\end{equation}
Evaluating this using the results in Fig~\ref{fig:combpp1}, we find that a detection of zero planets, $n<1$, corresponds to
\begin{equation}
\eta_{\rm LZ} < 0.18 \left(\frac{q}{10^{-3}}\right)^{-1} \left(\frac{\Delta\chi^2}{25}\right)^{1/2}.
\end{equation}

\subsection{PLANET EXCLUSION ZONES VS $a$(AU) AND $m(m_{JUP})$}
\begin{figure}
\centering
\begin{tabular}{c}
\psfig{file=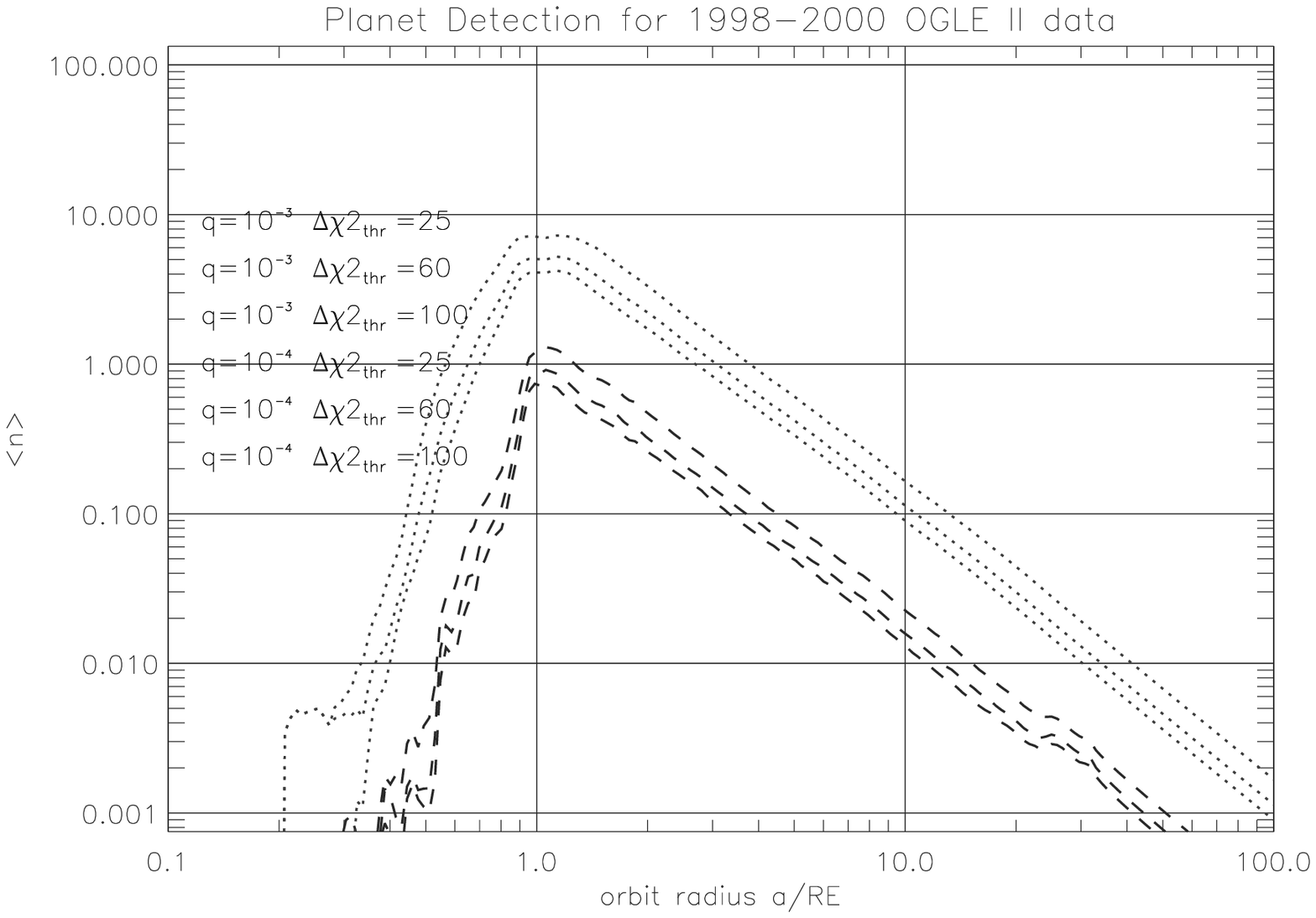,angle=0,width=8cm}
\end{tabular}
\caption[Total detection probability -vs- $a/R_{E}$]{\small Number of planet detections expected from our analysis of 145 OGLE microlensing events if all lens stars have $\eta=1$ planet with orbit radius $a$ in units of the Einstein ring radius $R_E$. Results shown are for two planet/star mass ratios, $q=10^{-3}$ (dotted) and $q=10^{-4}$ (dashed), and three detection thresholds $\Delta\chi^2=25,60,100$ (top to bottom).}
\protect\label{fig:combpp1}
\end{figure}

In Figure \ref{fig:combpp1} we have presented a general description of the total detection probability in terms of the mass ratios and the projected separation of the companion from the lens in units of the Einstein ring radius. Here we convert these to physical units using a similar method to that followed by \scite{Albrow01} to give a clearer indication of planetary constraints placed by the OGLE dataset. The results are shown in Fig~\ref{fig:combpp2}.

To convert to physical units we assume 
\begin{equation}
R_E = 1.9 \mbox{AU} \left( \frac{t_E}{45~{\rm d}} \right),     
\protect\label{eqn:re}
\end{equation}
and                                                
\begin{equation}                                        
M = 0.3 M_\odot \left( \frac{t_E}{45~{\rm d}} \right)^{1/2},
\protect\label{eqn:m}                        
\end{equation}
where 45d is the median event timescale for the OGLE events, 1.9 AU is the median Einstein ring radius, and 0.3 $M_\odot$ is the median lens mass. These scalings are consistent with a model in which most of the OGLE events arise from sources and lenses that are both galactic Bulge stars, with $D_d \sim 6$kpc, $D_s \sim 8$kpc.

This results in the combined detection probability for all OGLE events as a function of planetary orbital radius, as shown in Figure \ref{fig:combpp2}. Returning to Figure \ref{fig:combpp1}, we can see that both mass ratios peak at $a \approx R_E$. In particular, for  $q=10^{-3}$, the average total detection probability for the lensing zone is $P(\mbox{det}|a_1 < a < a_2) = 5.6$. From Figure \ref{fig:combpp2}, the average detection probability for the region $1\mbox{AU} < a < 4 \mbox{AU}$ is $P(\mbox{det}|1~\mbox{AU} < a < 4~\mbox{AU}) = 3.2$ for $q=10^{-3}$ and the probability peaks at around 3 AU. 

The curves in Fig~\ref{fig:combpp2} are broader and more symmetric than those in Fig~\ref{fig:combpp1}. This arises because the detection curves all peak at $a\approx R_E$, but each event has a different $R_E$ with the slower events corresponding on average to larger Einstein rings (eqn\ref{eqn:re}). This horizontal blurring in Fig~\ref{fig:combpp2} makes the curve wider than in Fig~\ref{fig:combpp1} and also reduces the expected number of detections to
\begin{equation}
\left<n\right> \approx 4 \eta \left(\frac{q}{10^{-3}}\right) \left(\frac{\Delta\chi^2}{25}\right)^{-1/2},
\end{equation}
for planets with $a\approx 3$AU, where the detection efficiency is highest.

For zero detections, $n<1$, the upper limit on the number of planets per lens star is                     
\begin{equation}                                                                
        \eta \la 0.23                                                           
        \left( \frac{q}{10^{-3}}\right)^{-1}                                    
        \left( \frac{\Delta\chi^2}{25}\right)^{1/2}                             
\ ,                                                                             
\end{equation}                                                                  
for planets with $a\approx 3$~AU, and                                            
\begin{equation}                                                                
        \eta_{1-4{\rm AU}} \la 0.31                                                   
        \left( \frac{q}{10^{-3}}\right)^{-1}                                    
        \left( \frac{\Delta\chi^2}{25}\right)^{1/2}                             
\ ,                                                                             
\end{equation}                                                                  
for planets uniformly distributed in $\log{a}$ over $a=1-4$~AU.
                                                 
The solid curves in Fig~\ref{fig:combpp1} show the expected number detections for              
Jupiter-mass planets.  With the star mass given by eqn~\ref{eqn:m},                    
the mass ratio for a planet of mass $m$ is                                      
\begin{equation}                                                                
        q = 3\times10^{-3}                                                      
        \left( \frac{t_E}{45{\rm d}}\right)^{-1/2}                              
        \left( \frac{m}{m_J}\right)                                             
        \ .                                                                     
\end{equation}                                                                  
The results for Jupiter-mass planets, $m=m_J$, are therefore found by a small linear extrapolation of those for $q = 10^{-3}$ (dotted) and  $q = 10^{-4}$ (dashed).                         
For zero detections, $n<1$, the upper limit on the number of planets per lens star is
\begin{equation}                                                                
        \eta \la 0.16                                                           
        \left( \frac{m}{m_J}\right)^{-1}                                        
        \left( \frac{\Delta\chi^2}{25}\right)^{1/2}                             
\ ,                                                                             
\end{equation}                                                                  
for planets with $a\approx 3$~AU, and                                            
\begin{equation}                                                                
        \eta_{1-4{\rm AU}} < 0.21                                                     
        \left( \frac{m}{m_J}\right)^{-1}                                    
        \left( \frac{\Delta\chi^2}{25}\right)^{1/2}                             
\ ,                                                                             
\end{equation}                                                                  
for planets uniformly distributed in $\log{a}$ over $a=1-4$~AU.

\section{UPPER LIMITS ON NUMBER OF PLANETS PER STAR}
Our analysis of the OGLE events provides evidence for $n$ planets, where $n \le 2$, since only 2 candidate planetary lens anomalies were identified, and some or both of these may be due to noise glitches rather than real planets.                                                
If we assume that all stars have $\eta$ planets with orbit radius $a$ and mass ratio $q$ uniformly distributed in log($\eta$), then the expected number of events is                                                                        
\begin{equation}
        <n> = \eta P(q,a),
\end{equation}
where                                                                                      
\begin{equation}
        P(a,q) = \sum_s P_s( \mbox{det}| q, a ).                                                    
\end{equation}

\begin{figure}
\centering
\begin{tabular}{c}
\psfig{file=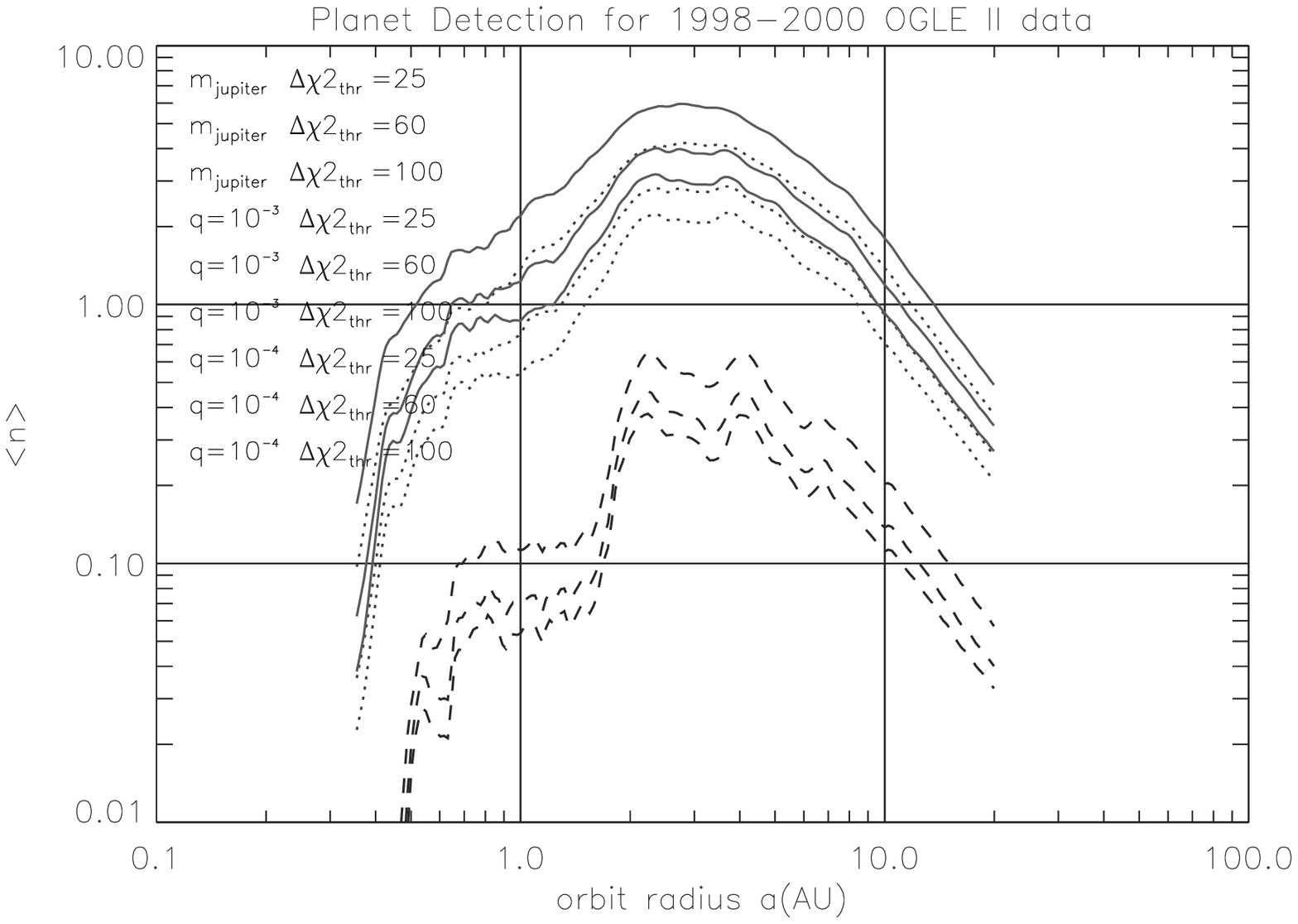,angle=0,width=8cm}
\end{tabular}
\caption[Total detection probability -vs- AU]{\small Number of planet detections expected from our analysis of 145 OGLE microlensing events if all lens stars have $\eta=1$ planet with orbit radius $a$ in units of the Einstein ring radius $R_E$. Results shown are for Jupiter mass planets (solid) and two planet/star mass ratios, $q=10^{-3}$ (dotted) and $q=10^{-4}$ (dashed), and three detection thresholds $\Delta\chi^2=25,60,100$ (top to bottom).}
\protect\label{fig:combpp2}
\end{figure}

\begin{figure}
\centering
\begin{tabular}{c}
\psfig{file=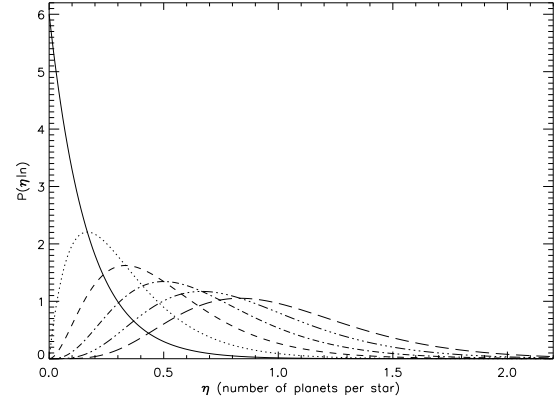,angle=0,width=8cm}
\end{tabular}
\caption[$P(\eta|n)$-vs-$\eta$]{\small $P(\eta|n)$-vs-$\eta$. Probability distribution of $\eta$ (the number of planets per star) for $n$=0,1,2,3,4,5 where $n$ is the number of detected planet anomalies in the dataset. $n$=0 is the leftmost curve.}
\protect\label{pfvn}
\end{figure}

From Bayes theorem, the probability distribution on $\eta$ given a measurement of $n$ is
\begin{equation}
P(\eta | n) = P(n | \eta) \  P(\eta)/P(n).
\end{equation}
Since we have no knowledge on the distribution of $P(\eta)$ we shall assume $P(\eta) \propto \eta^{-1}$. The last term in the equation,
\begin{equation}
P(n) = \int P(n | \eta) P(\eta) \mbox{d}\eta,
\end{equation}
is a normalization constant to ensure that $\int P(\eta | n) \mbox{d}\eta = 1$.
$P(n | \eta)$ is a likelihood function, i.e. the probability that we will detect $n$ planets given that there are $\eta$ per star. The number of detected planets $n$ is a Poisson random variable with expected value $<n> = \eta P(q,a) = E$. Then we have that
\begin{equation}
P(n | \eta) = \frac{E^{n} \mbox{exp}(-E)}{n!} \ .
\end{equation}
If the data hold no conclusive evidence of planets, i.e. $n=0$, then the constraint on $\eta$ is an exponential distribution:
\begin{equation}
P(\eta | n=0) = \frac{P(n=0 | \eta) \ P(\eta)}{\int_0^{\infty} P(n=0 | \eta) P(\eta) \mbox{d}\eta}.
\end{equation}
The general form of $P(\eta | n)$ can easily be calculated to be:
\begin{equation}
P(\eta | n) = \frac{f^n \ P(q,a)^{n+1} \ \mbox{exp}(-\eta \ P(q,a))}{n!}.
\end{equation}
This is shown on Figure \ref{pfvn}. 
If there are $n$ deviations which we assume are due to planets, then the expected number of planets per star of type $q,a$ is:
\begin{equation}
<\eta> = \int \eta P(\eta | n) \mbox{d}\eta = \frac{n}{P(q,a)}.
\label{eqn15}
\end{equation}
Under our assumptions and for a detection threshold value of $\Delta\chi^2 \geq 25$, we can state from the results presented in Figure \ref{fig:combpp2} that the OGLE dataset indicates that less than 21 ($n$) \%, where $n \le 2$, of the lens stars have Jupiters orbiting them at an orbital radius of $1 < a < 4$ AU.

Our result uses a smaller $\Delta\chi^2$ threshold than the results presented by the PLANET \cite{Gaudi02,Albrow01} collaboration. After searching for planetary signatures in 43 intensively monitored microlensing events, they concluded that less than 33\% of the $\sim 0.3 M_{\odot}$ stars that typically comprise the lens population have Jupiter-mass companions with semi-major axes in the range of $1 < a < 4$ AU. We must note however that their detection threshold value was very conservatively set at $\Delta\chi^2 \geq 60$. For comparison, we calculate the detection probability for the same region ($1.5 < a < 4$ AU) for $\Delta\chi^2 = 60$ to be $\sim 28$\%. This is lower than the limit imposed by PLANET (also note that our model assumptions lead to an {\it underestimate} of the true detection probability of the order of $\sim$13\%).

\section{SUMMARY}
We have analyzed three years (1998-2000) of OGLE observations of microlensing events to place limits on the abundance of `cool Jupiters'. We fitted a total of 145 events using a maximum likelihood fit that adjusts 6 parameters. We computed detection probability maps for each event using $\Delta\chi^2$ threshold values of 25, 60, 100. If we assume that all lenses are of the same star type and all have one Jupiter analogue at $a/R_{E} \sim 1$, then from the combined detection probability results we infer that a maximum of 7 planets with $q=10^{-3}$ should have been detected. Our selection criteria returned 5 candidate events for an assumed mass ratio of $q=10^{-3}$. Three of these were then confirmed to be due to other effects so only two remain as plausible candidates. Our results suggest that less than 14\% of the lens stars have `cool Jupiters' orbiting them at $a/R_{E} \sim 1$. We translate that in AU, recalculate the probability for Jupiter-mass planets, and from Figure \ref{fig:combpp2} we can state that the OGLE dataset indicates that less than 21 ($n$)\% of the lens star population has Jupiters orbiting them at an orbital radius of $1 < a < 4$ AU. $n \le 2$ is the number of planet anomaly candidates that are actually due to planets. We conclude that observing time is more efficiently allocated by observing many events with sampling intervals that produce non-overlapping detection zones than using intensive sampling on a few events.

\section{ACKNOWLEDGMENTS}
We thank the OGLE team for placing their data on these events in the public domain and in particular A. Udalski and P. Wozniak for useful feedback. Keith Horne was supported by a PPARC Sr. fellowship and by Beatrice Tinsley visiting Professorship at the University of Texas, Austin.

\bibliographystyle{mn}
\bibliography{iau_journals,master,Xplanets}

\appendix
\section{Excluded events}
We give a brief description of the events excluded from the PSPL fits and briefly discuss the possible nature of the deviation.

\underline{1998bul12:}
Having a baseline $I$ magnitude of 12.932, this event shows magnitude variations not consistent with the PSPL model. The magnitude starts increasing at JD$-2450000 \sim 920$ and reaches a maximum value of $\sim 12.67$mag at JD$-2450000 \sim$ 950 . Then it drops off but not following the standard Paczynski bell-shaped curve. The lightcurve observed could be that of a variable star.

\underline{1998bul28:}
This event displays a double peaked structure. It could be interpreted as a binary lens or a binary source. The PLANET collaboration has obtained data on this event and they find the best binary-lens model in the range $q=10^{-4}-10^{-2}$ has $\Delta\chi^2 \sim 19$ \cite{planetcons99}.

\underline{1998bul32:}
This event is sparsely monitored about the peak which slackens our interpretation of its likely nature. It exhibits a magnitude change of $\Delta I > 6$mag and clearly deviates from the PSPL model. We can make no definite claims about the cause of this variability.

\underline{1999bul11:}
This is a binary lens event with clear signs of caustic crossings. The amplification increases rapidly as the source crosses the first caustic, then falls off temporarily only to increase again as the source crosses the second caustic. The magnitude then returns to its baseline value.

\underline{1999bul17:} (also MACHO 99-BLG-28)
The data on this event are consistent with binary microlensing.

\underline{1999bul19:}
A superposition of two lightcurves resulting from binary source lensing can explain the effects seen in this lightcurve.

\underline{1999bul23:}
Anomalies expected from binary microlensing events are obvious in this lightcurve as well. The PLANET collaboration have published a paper presenting their fits to this event. The source star is a G/K sub-giant in the Galactic Bulge. Their best-fit parameters are a mass ratio $q \simeq 0.39$ and projected separation $d \simeq 2.42$ \cite{planet01}.

\underline{1999bul25:} (also MACHO 99-BLG-35)
Based only on the OGLE public available data, it remains unclear whether this event is a binary lens or a binary source. The deviations around the time of peak amplification are short lived and can be accounted for by both models. However, extra data obtained by the PLANET collaboration, but unavailable to the community, strongly suggest that this event is caused by a binary lens.

\underline{1999bul28:}
A very short timescale event with $t_{E}$ (radius) of 5.972 days. Sparsely monitored about the peak, it is not obvious what the slight deviations from the PSPL model are due to.

\underline{1999bul29:}
The PLANET collaboration excludes event 1998BUL29 from their analysis since they find that a point-lens finite-source model fits their data better. We exclude it too.

\underline{1999bul32:}
There seem to be a number of effects present in this lightcurve. Since this is is a long timescale event with $t_{E} \simeq 155$ (radius) we expect the parallax effect due to the Earths motion to be detectable. The event also seems to be severely blended.

\underline{1999bul40:}
A faint event with baseline magnitude $I$=19.780. OGLE data do not help distinguish what causes the observed deviation. PLANET have unpublished data on this event and claim this is another binary lensing event.

\underline{1999bul42:}
For 1999BUL42 the binary nature of the lens is seen in the lightcurve. The deviation lasts for $\sim 21$ days. No fits are available for the data.

\underline{2000bul28:}
Recently reduced PLANET data from two observing sites appear to suggest that this event is due to multiple lens microlensing. Two caustic crossings may have occurred at JD-2450000=1677 and JD-2450000=1682.

\underline{2000bul38:}
This is a caustic crossing binary lens event.

\underline{2000bul43:}
A long timescale event with $t_{E} \simeq 174$ (radius) days. The event is observed only during the increasing amplification phase and the parallax effect due to the Earths orbital motion is obvious in the lightcurve.

\underline{2000bul46:}
A binary lens event. 

\bsp
\end{document}